\documentclass{aa}

\usepackage{graphicx}
\usepackage{txfonts}

\usepackage[colorlinks=true,urlcolor=blue,citecolor=red,linkcolor=red,bookmarks=true]{hyperref}

\def\lb#1{{\protect\linebreak[#1]}}
\def\swift{{\tt http://\lb{2}www.boulder.\lb{2}swri.edu/\lb{2}{\~{ }}hal/\lb{2}swift.html}}
\def\httppisa{{\tt https://\lb{2}newton.\lb{2}spacedys.\lb{2}com/\lb{2}astdys/}}
\def\httpdamit{{\tt http://\lb{2}astro.troja.\lb{2}mff.cuni.cz/\lb{2}projects/\lb{2}damit/}}
\def\httplcdb{{\tt http://\lb{2}www.minorplanet.\lb{2}info/\lb{2}lightcurvedatabase.html}}

\newcommand{\gammaSI}{{\,J\,m$^{-2}$\,s$^{-0.5}$\,K$^{-1}$}}

\begin{document}

\title{(208) Lacrimosa: A case that missed the Slivan state?}

\author{D. Vokrouhlick\'y\inst{1} \and
        J. \v{D}urech\inst{1}     \and
        J. Hanu\v{s}\inst{1}      \and
        M. Ferrais\inst{2}        \and
        E. Jehin\inst{3}          \and
        Z. Benkhaldoun\inst{4}
        }

\titlerunning{(208) Lacrimosa: A case that missed the Slivan state?}
\authorrunning{Vokrouhlick\'y et~al.}

\institute{Institute of Astronomy, Charles University, V~Hole\v{s}ovi\v{c}k\'ach 2,
           CZ-180~00 Prague 8, Czech Republic \\
           \email{vokrouhl@cesnet.cz,durech@sirrah.troja.mff.cuni.cz}  \and
           Aix-Marseille Universit\'e, Laboratoire d'Astrophysique de Marseille,
           38, rue Fr\'ed\'eric Joliot-Curie, F-13388 Marseille, France \and
           Space Sciences, Technologies and Astrophysics Research Institute, Universit{\'e} de
           Li{\`e}ge, All{\'e}e du 6 Ao{\^u}t 17, B-4000 Li{\`e}ge, Belgium \and
           Oukaimeden Observatory, High Energy Physics and Astrophysics Laboratory,
           Cadi Ayyad University, Marrakech, Morocco
           }

\date{Received: \today ; accepted: ???}

\abstract
{The largest asteroids in the Koronis family (sizes $\geq 25$~km) have very peculiar
 rotation state properties, with the retrograde- and prograde-rotating objects being
 distinctly different. A recent re-analysis of observations suggests that one of the asteroids
 formerly thought to be retrograde-rotating, 208~Lacrimosa, in reality exhibits
 prograde rotation, yet other properties of this object are discrepant with other members this group.}
{We seek to understand whether the new spin solution of Lacrimosa invalidates the previously
 proposed model of the Koronis large members or simply reveals more possibilities
 for the long-term evolutionary paths, including some that have not yet been explored.}
{We obtained additional photometric observations of Lacrimosa, and included thermal
 and occultation data to verify its new spin solution. We also conducted a more detailed
 theoretical analysis of the long-term spin evolution to understand the discrepancy with
 respect to the other prograde-rotating large Koronis members.}
{We confirm and substantiate the previously suggested prograde rotation of Lacrimosa. Its
 spin vector has an ecliptic longitude and latitude of $(\lambda,\beta)=(15^\circ \pm 2^\circ,
 67^\circ\pm 2^\circ)$ and a sidereal rotation period $P=14.085734\pm 0.000007$~hr. The
 thermal and occultation data allow us to calibrate a volume equivalent size of
 $D=44\pm 2$~km of Lacrimosa. The observations also constrain the shape model relatively well.
 Assuming uniform density, the dynamical ellipticity is $\Delta=0.35\pm 0.05$. Unlike other large
 prograde-rotating Koronis  members, Lacrimosa spin is not captured in the Slivan
 state. We propose that Lacrimosa differed from this group in that it had initially slightly
 larger obliquity and longer rotation period. With those parameters, it jumped over the
 Slivan state instead of being captured and slowly evolved into the present spin
 configuration. In the future, it is likely to be captured in the
 Slivan state corresponding to the proper (instead of forced) mode of the orbital plane
 precession in the inertial space.}
{}

\keywords{Celestial mechanics -- Minor planets, asteroids: general}

\maketitle


\section{Introduction}
Modern automated surveys have revolutionized our knowledge of the near and far universe 
in many respects. When complemented with the dedicated efforts of specific individual
projects, sometimes also supported by observations of amateur astronomers, 
our current knowledge largely surpasses what we had two or three decades ago. Consider,
as an example, well-calibrated photometric observations, which are now available for 
a sufficient period of time for determination of the rotation state of the minor bodies
in the Solar System. As of now, we have information about 
rotation periods for tens of thousands of asteroids in the near-Earth and main-belt 
populations \citep[e.g.,][updated as of October 2020 on \httplcdb]{lcdb}. For several
thousand among them we have additional information about the orientation of their spin
axis and basic shape parameters \citep[e.g.,][and updates on \httpdamit]{damit}. These
numbers have grown so large that they enable population-scale studies, rather than simple
analyses of individual objects \citep[e.g.,][]{detal15}.

One of the first examples of an interesting result from this tremendous progress was the
unexpected discovery 
of the nonrandom distribution of rotation states among large members in the Koronis family 
by \citet{slivan02} \citep[see also further details in][]{setal2003,setal2008,setal2009}. At
first glance, the fact that the Koronis family is $2$ to $3$~Gyr old, and formed
likely by a super-catastrophic collision \citep[e.g.,][]{netal15}, would lead us to
expect a random
distribution of the rotation states of its large members. In particular, rotation
periods were expected to be consistent with a Maxwellian distribution and the direction
of rotation poles isotropic in space. In stark contrast, observations reported by
\citet{slivan02} told an entirely different story. Of the ten $D\geq 25$~km objects,
six were found to rotate retrograde, (i) having either slow or fast
rotation (periods $P\leq 4.63$~hr or $P\geq 13.06$~hr), and (ii) rotation poles
pushed toward the south ecliptic pole (obliquities $\varepsilon\geq 154^\circ$).
Even more puzzling was the set of four prograde-rotating objects (i)
whose rotation periods were all within a rather tight interval of values ($7.5<
P<9.5$~hr), and (ii) whose rotation poles were near to parallel in the inertial space
(within about $50^\circ$ cone), all having obliquity $\simeq 45^\circ$.

All these astonishing findings were soon reconciled with Koronis long-term history
by \citet{vetal2003}. These authors demonstrated that the missing key element in the
pre-2000 thinking was the Yarkovsky-O'Keefe-Radzievskii-Paddack (YORP) effect,
reintroduced into the planetary studies by \citet{rub2000} \citep[see also][for
an overview of its history and
recent status]{vetal15}. YORP is a weak, nonconservative torque capable, in the
long term, of either accelerating or decelerating the rotation rate, and at the same
time tilting the spin axis toward extremal values of the obliquity ($0^\circ$ or
$180^\circ$). \citet{vetal2003} noted that for their sizes and heliocentric distance,
asteroids in the Koronis family, whose rotation was initially retrograde,
would roughly complete such an evolution toward the asymptotic period and obliquity
values just within its expected age. This would readily explain the group of
retrograde rotators observed by \citet{slivan02}. The group of prograde rotators
were more difficult to explain, because the pattern reported by \citet{slivan02} was
not symmetric. Here the additional key element is the intriguing interplay between
the effects of gravitational torque due to the Sun and  the motion of the asteroids'
heliocentric orbital plane. Regular precession due to the former phenomenon may enter
into a resonance with precession of the latter (see Appendix~\ref{appa1}). A
possibility for such secular spin orbit resonance exists
only for prograde-rotating bodies. Assuming elongated shapes compatible with
light curve observations, and periods of $\simeq 8$~hr of the prograde group of large
Koronis objects, the resonance in question would be located at about $40^\circ-
50^\circ$ obliquity. Importantly, near this
value, the YORP evolution of the rotation rate temporarily stalls \citep[e.g.,][]
{rub2000,vc2002,cv2004}. This means that, while still evolving by YORP, the
large Koronis prograde-rotating asteroids may spend giga years near such a temporary
state once captured in the resonance. \citet{vetal2003} also proved that
when YORP previously brought the spin towards a small obliquity state while
decelerating the rotation rate, the capture into the resonance must occur.
Finally, the apparently most puzzling observation, namely spin parallelism in
the inertial space, is also readily explained by the above-mentioned spin
orbit resonance. This is because the particular precession mode of the heliocentric
orbital plane that resonates with regular precession of the Koronis asteroids
is forced by the current configuration of giant planets, in particular the direction
of the orbital node of Saturn. As a result, there is no mysterious direction in the inertial
space due to distant cosmic objects that would attract rotation poles of Koronis
members, but simply the resonance stationary point ---about which they librate---
has a specific direction related to the configuration of the orbital planes of the
giant planets. In order
to pay tribute to the painstaking observational work of Steve Slivan that brought to life
all these elegant theoretical concepts, \citet{vetal2003} proposed naming the
spin-orbit resonant state, in which the prograde-rotating Koronis members are
locked, ``the Slivan state''.

Focusing still on the sample of $D\geq 25$~km objects in the Koronis family, we
note that the follow-up work of \citet{setal2009} reported a fifth member in the
Slivan state with very similar rotation parameters to the other prograde-rotating
Koronis members, namely (462)~Eryphila, further strengthening the story. However,
these latter authors also found evidence of a first stray prograde-rotating object with somewhat
divergent parameters, namely (263)~Dresda. In particular, Dresda's obliquity was
found to be only $\simeq 15^\circ$ and its rotation period $\simeq 16.8$~hr (we note,
however, that \citet{hetal2016} corrected this solution, bringing
the pole closer to the original Slivan group with an obliquity of $\simeq 35^\circ$).

Information about the rotation state of smaller Koronis family members has been
provided by the analysis of data from all-sky surveys from the past
decade or so \citep[we purposely omit the interesting case of (832)~Karin][saving
it for a future detailed study]{karin2012}. \citet{hetal2013}, followed with
\citet{hetal2016} and \citet{detal2019}, determined the spin states of thousands of
asteroids, among them also 13 $D\leq 25$~km members of the Koronis family.
Many of them, especially among the retrograde rotators, follow the trends first
observed by \citet{slivan02}, but some do not. We comment on the implications of this
below. However, one of these new spin models concerned a $D\simeq 45$~km
Koronis member, asteroid (208)~Lacrimosa, which was included in the original
study of \citet{slivan02} and belonged to the retrograde group of bodies with
rather long rotation periods.%
\footnote{Lacrimosa belongs to the largest members in the Koronis family. In fact,
 \citet{metal13} opted to call the cluster ``Lacrimosa family'', which proved
 unsuitable because of a long tradition and history of the Koronis family since
 the pioneering work of \citet{hira18}.}
\citet{detal2019} pointed out that this solution was incorrect. While the
rotation period of $\simeq 14.086$~hr was in agreement with their findings, the rotation
pole in their
solution moved to the prograde group with two possible solutions for the ecliptic 
longitude $\lambda$ and latitude $\beta$, namely $(\lambda,\beta)_1 = (16^\circ,
60^\circ)$ or $(\lambda,\beta)_2=(202^\circ,61^\circ)$.
Curiously, the first pole solution would fit rather well with the originally
reported group of Koronis prograde rotators in the Slivan state, but the rotation
period is longer.

This new solution for one of the original Slivan targets prompted us to
re-evaluate the situation and see if the above-outlined story of \citet{vetal2003}
still holds. Before we deal with this primary goal (in Sect.~\ref{results}), we
first present the current rotation-state solution for (208)~Lacrimosa in more
detail (Sect.~\ref{data}). In particular, to confirm the stability of the solution, we
obtained new observations during the last Lacrimosa opposition and added them to the
full observational record for this asteroid. Additionally, we included sparse photometric
observations from numerous sky surveys and stellar
occultations from two different epochs. We then conducted a
numerical exploration of its short- and long-term evolution (Sect.~\ref{results}).
Some details of the mathematical methods and numerical tools are summarized in 
Appendix~A. Basic information about our new observations of Lacrimosa are
given in Appendix~B. Our best-fitting model is compared with all available 
observations in Appendix~C.

\section{Rotation state of (208) Lacrimosa} \label{data}
As mentioned above, the spin state of (208) Lacrimosa published by \cite{detal2019} was
different from that in \cite{slivan02} and \citet{setal2003}. To ensure that the new
pole solution is correct, we repeated the light-curve inversion with a much larger
dataset. We collected all available light curves \citep{bin1987,sb1996,s2014} and
sparse photometry from Gaia DR2 \citep{gaia2018}, ASAS-SN 
\citep[All-Sky Automated Survey for Supernovae;][]{Shappee2014b, Kochanek2017},
ATLAS \citep[Asteroid Terrestrial-impact Last Alert System;][]{tetal2018}, and
United States Naval Observatory (USNO) and Catalina observatories downloaded
from Minor Planet Center (MPC). We also
carried out new photometric observations of Lacrimosa with TRAPPIST-South and TRAPPIST-North
telescopes in March and June 2020 \citep[e.g.,][]{jeh2011}. Some technical details of 
these new observations and their reduction methods are given in Appendix~B. All 
photometric data used for the inversion are listed in Table~\ref{tab:aspect}, and 
their comparison with the best-fitting model is shown in Appendix~C.
\begin{table*}[t]
 \caption{\label{tab:aspect}    
  Aspect data for available observations of (208) Lacrimosa. The table lists its distance
  from the Sun $r$ and from the Earth $\Delta$, the solar phase angle $\alpha$,
  its geocentric ecliptic coordinates $(\lambda, \beta)$, and the
  observatory or source of data. Our new observations taken in March and June 2020 were
  made as part of the TRAPPIST survey. Sparse-in-time photometry is listed at the bottom
  of the table and covers a wide range of geometries. The data come from Gaia Data
  Release 2, All-Sky Automated Survey for Supernovae, Asteroid Terrestrial-impact
  Last Alert System (cyan and orange filters), the US Naval Observatory, and the Catalina
  Sky Survey.}
 \centering
  \begin{tabular}{cccrrrl}
   \hline \hline
   Date & $r$ & $\Delta$ & $\alpha\phantom{g}$ & \multicolumn{1}{c}{$\lambda$} &
    \multicolumn{1}{c}{$\beta$} & Obs. \\
   & [au] & [au] & [deg] & \multicolumn{1}{c}{[deg]} & \multicolumn{1}{c}{[deg]} & \\
   \hline
   & & & & & & \\ [-4pt]
  \multicolumn{7}{c}{\em --Dense photometry--} \\ [1pt]
  1985 02 15.4  & 2.872  & 2.158  & 15.8   & 199.0  & $ 0.1$ & \cite{bin1987} \\
  1989 02 03.2  & 2.860  & 1.991  & 11.2   & 100.4  & $ 2.4$ & \cite{sb1996} \\
  1989 02 04.2  & 2.860  & 1.998  & 11.5   & 100.3  & $ 2.4$ & \cite{sb1996} \\
  1990 03 31.3  & 2.887  & 1.899  &  3.6   & 200.9  & $-0.6$ & \cite{sb1996} \\
  1992 10 24.2  & 2.897  & 1.919  &  4.4   &  18.3  & $ 0.8$ & \cite{sb1996} \\
  1992 11 16.2  & 2.895  & 2.061  & 12.5   &  14.8  & $ 0.9$ & \cite{sb1996} \\
  1992 11 21.1  & 2.894  & 2.108  & 13.9   &  14.4  & $ 1.0$ & \cite{sb1996} \\
  1994 01 09.3  & 2.864  & 1.883  &  2.0   & 114.0  & $ 2.5$ & \cite{sb1996} \\
  1994 01 10.3  & 2.864  & 1.882  &  1.6   & 113.8  & $ 2.5$ & \cite{sb1996} \\
  1994 01 11.2  & 2.864  & 1.881  &  1.3   & 113.6  & $ 2.5$ & \cite{sb1996} \\
  1994 01 16.2  & 2.863  & 1.881  &  1.5   & 112.5  & $ 2.5$ & \cite{sb1996} \\
  1994 01 19.2  & 2.863  & 1.885  &  2.6   & 111.8  & $ 2.5$ & \cite{sb1996} \\
  2014 01 14.3  & 2.857  & 1.990  & 11.2   & 148.2  & $ 1.8$ & \cite{s2014} \\
  2014 01 14.5  & 2.857  & 1.989  & 11.1   & 148.1  & $ 1.8$ & \cite{s2014} \\
  2014 01 15.3  & 2.857  & 1.982  & 10.9   & 148.0  & $ 1.8$ & \cite{s2014} \\
  2014 01 15.5  & 2.857  & 1.981  & 10.8   & 148.0  & $ 1.8$ & \cite{s2014} \\
  2014 01 16.3  & 2.857  & 1.975  & 10.5   & 147.9  & $ 1.8$ & \cite{s2014} \\
  2014 01 16.5  & 2.857  & 1.973  & 10.5   & 147.9  & $ 1.8$ & \cite{s2014} \\
  2020 03 07.3  & 2.906  & 2.562  & 19.6   & 246.7  & $-1.3$ & TRAPPIST-South \\
  2020 03 08.2  & 2.907  & 2.550  & 19.6   & 246.9  & $-1.4$ & TRAPPIST-North \\
  2020 03 09.1  & 2.907  & 2.537  & 19.5   & 247.0  & $-1.4$ & TRAPPIST-North \\
  2020 03 09.2  & 2.907  & 2.536  & 19.5   & 247.0  & $-1.4$ & TRAPPIST-South \\
  2020 03 10.2  & 2.907  & 2.523  & 19.5   & 247.2  & $-1.4$ & TRAPPIST-North \\
  2020 03 10.3  & 2.907  & 2.522  & 19.5   & 247.2  & $-1.4$ & TRAPPIST-South \\
  2020 03 11.2  & 2.907  & 2.510  & 19.4   & 247.3  & $-1.4$ & TRAPPIST-North \\
  2020 03 11.3  & 2.907  & 2.508  & 19.4   & 247.3  & $-1.4$ & TRAPPIST-South \\
  2020 03 12.3  & 2.907  & 2.495  & 19.3   & 247.4  & $-1.4$ & TRAPPIST-South \\
  2020 03 19.3  & 2.908  & 2.403  & 18.7   & 248.2  & $-1.5$ & TRAPPIST-South \\
  2020 06 02.0  & 2.915  & 1.914  &  4.0   & 240.3  & $-2.3$ & TRAPPIST-North \\
  2020 06 03.0  & 2.915  & 1.917  &  4.4   & 240.2  & $-2.3$ & TRAPPIST-North \\
  2020 06 07.0  & 2.916  & 1.931  &  5.9   & 239.4  & $-2.3$ & TRAPPIST-North \\
  2020 06 11.0  & 2.916  & 1.949  &  7.4   & 238.7  & $-2.3$ & TRAPPIST-North \\
  2020 06 14.9  & 2.916  & 1.970  &  8.8   & 238.0  & $-2.3$ & TRAPPIST-North \\
  2020 06 18.0  & 2.917  & 1.989  &  9.9   & 237.6  & $-2.3$ & TRAPPIST-North \\ [2pt]
  \multicolumn{7}{c}{\em --Sparse photometry--} \\ [1pt]
  2015/01--2016/05  &    &   &      &      &                          & Gaia DR2 \\
  2012/10--2018/11 &     &   &      &  &                              & ASAS-SN \\
  2015/08--2018/02  &     &   &       &      &                        & ATLAS c \\
  2015/08--2018/09 &  &  &   &   &                                    & ATLAS o \\
  1998/11--2009/02 &  &  & &   &                                      & USNO \\
  2003/11--2016/09  & &  &    &   &                                   & Catalina \\
  \hline
 \end{tabular}
\end{table*}

\begin{figure}[t]
 \begin{center} 
 \includegraphics[width=0.49\textwidth, trim=2cm 17cm 1cm 3cm, clip]{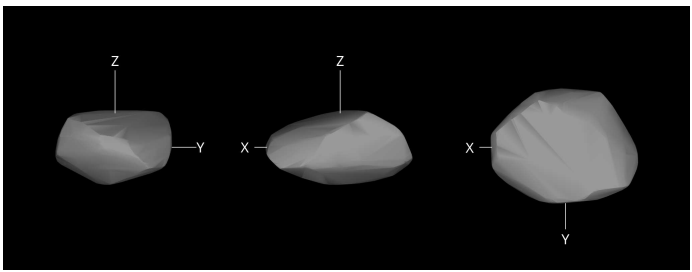}
 \end{center}
 \caption{\label{fig:shape}
  Shape model of Lacrimosa for pole direction $(15^\circ, 67^\circ)$ shown
  from equatorial level (left and center, $90\degr$ apart) and pole-on (right).}
\end{figure}

From light curves and sparse photometry, we reconstructed two convex shape models with
the inversion method of \cite{kaa2001}. One of the models (shown in Fig.~\ref{fig:shape})
has the pole direction $(15^\circ \pm 2^\circ, 67^\circ \pm 2^\circ)$ in ecliptic longitude and latitude
and its rotation period is $P = 14.085734\pm 0.000007$\,hr. The second model has the same
rotational period, and its pole direction is $(204^\circ,  68^\circ)$. Both models provide
the same RMS fit of the data. Because Lacrimosa's orbital inclination to the ecliptic is
only $1.7^\circ$, the viewing and illumination geometry of observations is always limited
to the ecliptic plane. For that reason, disk-integrated photometry can never distinguish
between these two symmetric pole solutions, which have the same ecliptic latitude and
ecliptic longitudes that are $180^\circ$ apart \citep{kl2006}. The uncertainty on spin parameters
given above was estimated using a bootstrap approach. We created $1000$ bootstrapped data sets by randomly
resampling light curves and sparse data points and repeated the light-curve inversion. For
each resampling, the inversion algorithm converged to a slightly different set of
parameters. Their standard deviation served as an estimate of their uncertainties.

The new spin solution is different from that derived by \cite{setal2003}, which was also
used in the original spin-clustering paper by \cite{slivan02}. These latter authors derived a rotation
period of $14.07692 \pm 0.00002$\,h and a retrograde pole solution. Their result was
based on a limited data set (see Table~\ref{tab:aspect}), and was  apparently only one 
of several local minima in the parameter space. Indeed, this weakness was noted already
by \cite{setal2003}, who mentioned: {\it ``The pole results for Lacrimosa
are preliminary and should be checked by further observations; especially needed are a
good single-apparition solar phase function and complete light curves at unobserved or
incompletely observed aspect longitudes.''} Our new analysis with a much larger data set
shows that the correct sidereal rotation period is slightly different from that of
\cite{setal2003}. Interestingly, this small discrepancy in periods leads to a dramatic difference
in spin axis directions, namely the change from retrograde to prograde rotation.

Our new model is also consistent with thermal infrared (IR) data from for IRAS, Akari, and WISE
observatories compiled in the Small Bodies: Near and Far Database \citep[SBNAF,][]{setal2020},
from where we downloaded processed fluxes. We used the approach of \cite{detal2017} and
reconstructed a model of Lacrimosa from its light curves combined with thermal data. There
were different possibilities for thermophysical parameters that gave similar fits to data,
one of them having thermal inertia $\Gamma = 30$\gammaSI, geometric albedo $p_V = 0.20$,
and volume-equivalent diameter $D = 44$~km. Its pole direction of $(13^\circ, 70^\circ)$ is
close to the value based on photometry alone. Because thermal data were also acquired at
plane-restricted geometries, the same symmetry applies here, and both pole directions are
equally good in fitting thermal data. Our solution utilizing thermal data is therefore
very close to that obtained by \citet{metal11}, who obtained $D=45.0\pm 4.6$~km and 
$p_V=0.168\pm 0.055$.
\begin{figure*}[t]
 \begin{center}
  \includegraphics[width=0.46\textwidth]{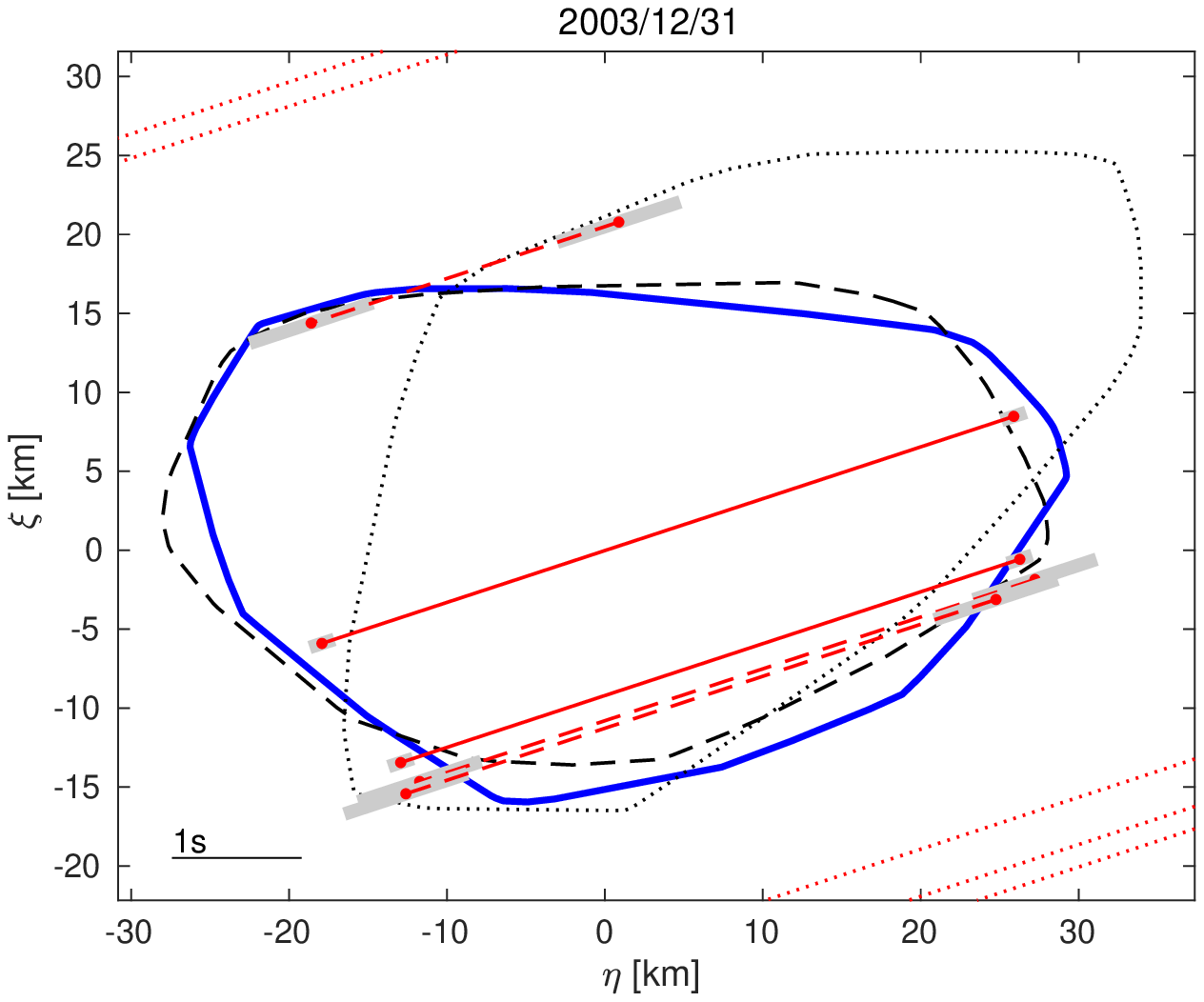}
  \includegraphics[width=0.46\textwidth]{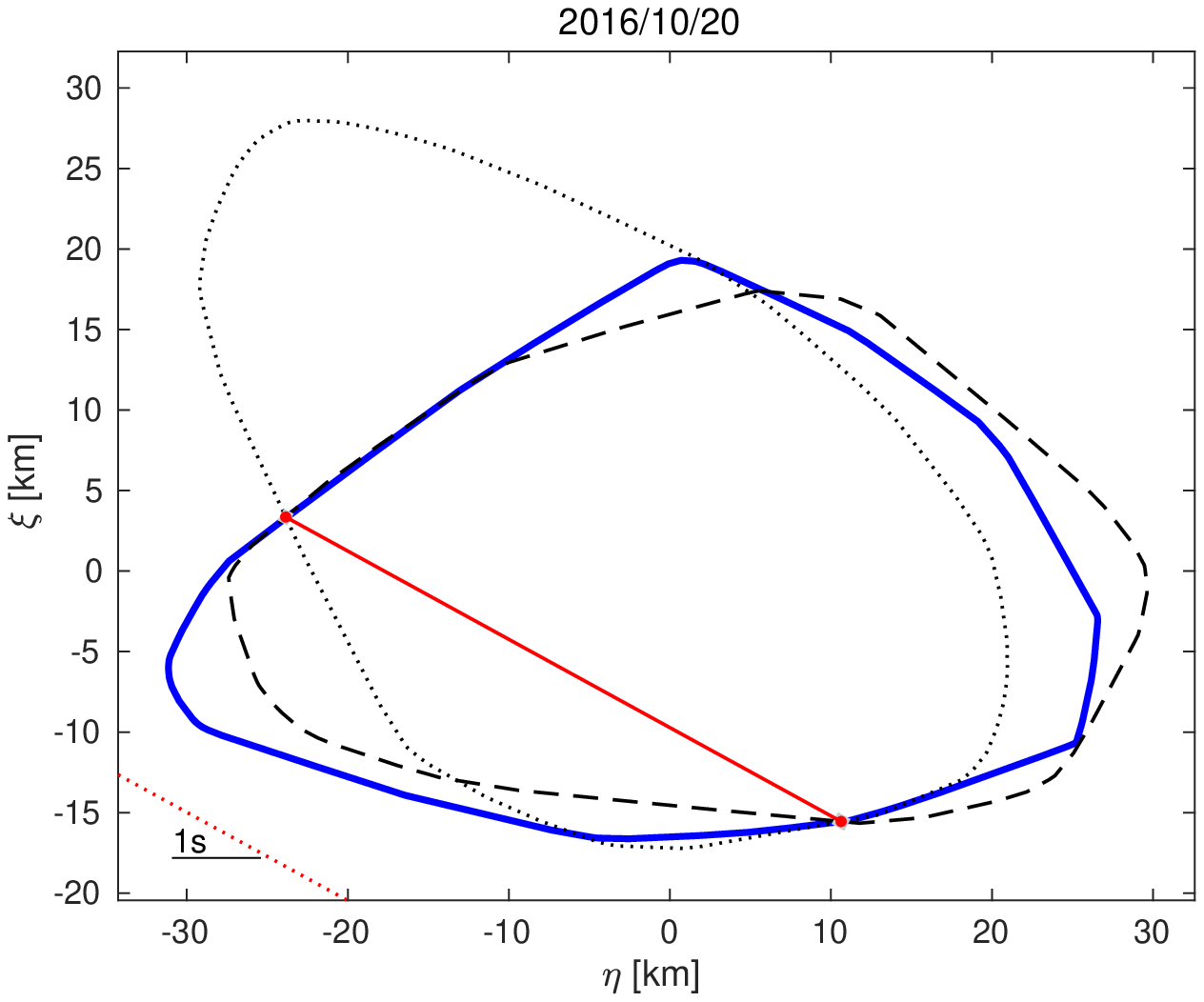}
 \end{center}
 \caption{\label{fig:occ_fit}
  Projections of two occultations from December 31, 2003
  (left), and October~20, 2016 (right). Individual observations are
  shown as straight red lines. Solid lines are photoelectric observations, dashed
  are visual observations, and dotted are negative observations. Timing errors are
  displayed as gray strips. The blue solid silhouette is that of the best-fit model, the
  dotted silhouette is of the second pole solution, and the dashed silhouette is of the
  IR-based shape model without any scaling. North is up, west to the right.}
\end{figure*}

Finally, there are two stellar occultations by Lacrimosa observed in 2003 and 2016
\citep{hetal2020}. We computed the orientation of our two models for the time
of occultations, computed the projected silhouettes, and scaled and shifted
the shape models to get the best agreement between the silhouettes and the
occultation chords \citep[for details, see][]{detal2011}. Because there were
no timing errors reported for the 2003 occultation, we assumed errors of 0.1
and 0.5\,s for photoelectric and visual observations, respectively. Only one
positive chord was observed during the occultation in 2016, and so the only constraint
comes from the 2003 occultation. The results are shown in Fig.~\ref{fig:occ_fit}.
Although the number of chords is not sufficient for any high-fidelity work, the
first pole solution $(15^\circ, 67^\circ)$ fits the occultation data better than
the second one with pole direction $(204^\circ, 68^\circ)$. The volume-equivalent
diameter is $44 \pm 2$\,km for the first model; this diameter is  $46 \pm 3$\,km
for the second model with much worse formal fit. For comparison, we also show
a silhouette of the shape model derived by simultaneous inversion of optical
and thermal data.

An important take-away experience from our analysis of Lacrimosa can be 
summarized as follows.    
Although further photometric observations can refine the shape model and
increase the accuracy of spin parameters, the pole ambiguity cannot be avoided by any
amount of disk-integrated data. The only way to distinguish between the two
spin axis directions is through disk-resolved data. For example, a well-observed
occultation would enable us to confirm that the $(15^\circ, 67^\circ)$ pole is
the correct one. Moreover, it could also help us to constrain the shape more tightly,
namely its dynamical ellipticity $\Delta$, which is  discussed in the following sections.
Nevertheless, because
the available occultation data already now favor this first photometric solution
of the pole of Lacrimosa, we consider it a viable solution in what follows.

\subsection{YORP torques for (208) Lacrimosa} \label{yorp}
\citet{vetal2003} pointed out that modeling of the very long-term evolution of
Koronis asteroid spin states requires inclusion of the YORP effect in the
dynamical model. We therefore need to estimate its strength. This task is quite
troublesome if high precision is required \citep[such as needed for comparison with
YORP detections on small near-Earth asteroids; see discussion in][]{vetal15}
but this is not the case here. Our goal is to simply characterize
the possible evolution of Lacrimosa's spins state that would result in its current
value. It is not our ambition, and it is not even possible, to hope for any
determinism in this task. Therefore, it is adequate to estimate the YORP effect
within a factor of a few in accuracy for our purposes.
\begin{figure*}[t]
 \begin{center} 
 \includegraphics[width=0.9\textwidth]{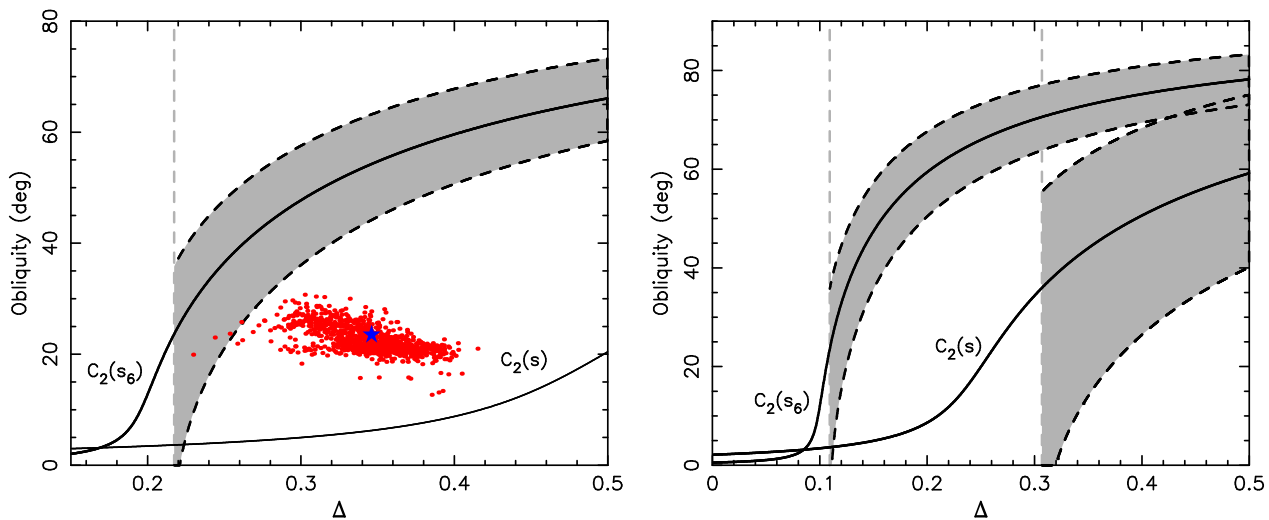}
 \end{center}
 \caption{\label{f1}
  Obliquity $\varepsilon_2$ of the Cassini state 2 as a function of the
  dynamical ellipticity $\Delta$ (see Eq.~\ref{flat}). Orbital parameters of
  (208) Lacrimosa are assumed. Left panel: Nominal rotation period $P=14.085734$~hr
  of (208)~Lacrimosa used. The solid line labeled $C_2(s_6)$ provides $\varepsilon_2$
  for the $s_6$ (forced) frequency mode of the nodal precession. The spin-orbit
  resonance onsets for $\Delta$ are denoted by the light-gray dashed line (transition
  determined by the Eq.~\ref{kappacrit} condition); beyond this value the Cassini
  state~2 becomes an equilibrium point of the spin-orbit resonance, whose maximum
  extension in obliquity is shown by the gray area. Solid line labeled $C_2(s)$ provides
  $\varepsilon_2$ for the $s$ (proper) mode of the nodal precession. Here  the
  spin-orbit resonance does not exist. Red symbols show obliquity and $\Delta$ values
  for a little less than 1000 solutions for (208) Lacrimosa from the bootstrap
  method discussed in Section~\ref{data} and using only the optical light-curve 
  observations. The blue star is the nominal, best-fit solution.
  Right panel: Same as in the left panel,
  but now for a hypothetical, longer rotation period of $P=28$~hr. Now the spin-orbit
  resonance exists beyond some critical $\Delta$ value for both frequencies $s_6$
  and $s$.}
\end{figure*}

We used the zero thermal conductivity approach of \citet{vc2002}, adopted the best-fit,
scale-calibrated model of Lacrimosa outlined above (volume-size corresponding to a
spherical body of diameter $\simeq 44$~km), and assumed a bulk density of $2$
g~cm$^{-3}$. With the parameters of the present spin state, and the heliocentric
orbit, we obtained: (i) the rate of change of the rotational frequency $\omega$
equal to $d\omega/dt \simeq -2.98\times 10^{-8}$ s$^{-1}$~Myr$^{-1}$, and (ii) the
rate of change of the obliquity $\varepsilon$ equal to $d\varepsilon/dt \simeq -0.014$
deg~Myr$^{-1}$ (we did not need to compute the YORP effect on ecliptic longitude,
because this contribution is much smaller than the precession due to solar gravitational
torque).
Both $\omega$ and $\varepsilon$ are thus predicted to decrease at this moment.
The current value of the doubling timescale \citep[e.g.,][]{rub2000} therefore 
reads $|\omega / (d\omega/dt)| \simeq 4.16$~Gyr. Another way of illustrating 
the YORP effect is to translate $d\omega/dt$ to the present-day rate of change of
the rotation period $P$. If this value is conserved, $P$  will increase by
$\simeq 3.4$~hr in the next gigayear. In reality the effect is even larger, because 
$dP/dt \propto P^2$ for an approximately constant $d\omega/dt$. As $P$ increases,
the rate $dP/dt$ therefore accelerates. The take-away message is that the YORP
effect is indeed fully capable of significantly changing Lacrimosa's rotation period
on a timescale of 1  Gyr,  which is comparable to the  age of the Koronis family.

To enable efficient long-term propagation of spin state with the YORP torques,
we also precomputed $d\omega/dt$ and $d\varepsilon/dt$ values for the
dense grid in obliquity (using $2^\circ$ step). In our simulations described in
Section~\ref{ev2} we simply interpolated among these values to obtain $d\omega/dt$
and $d\varepsilon/dt$ for an arbitrary obliquity (see Appendix~\ref{appa1}).

\section{Theory} \label{results}
The analysis of observations in the previous section provides parameters of the
rotation state at the current epoch. It assumes the spin orientation and
sidereal rotation rate are constant, at least over the few decades
covered by the data. Given the measurement accuracy, this is a justifiable assumption.
However, over a longer period of time all rotation-state parameters evolve. Here
we pay attention to secular effects, namely those with characteristic timescale
longer than the sidereal rotation period of the asteroid and its orbital period about
the Sun. We first characterize short-term secular effects (1 Myr timescale; 
Section~\ref{ev1}). This initial step is important for two reasons. First, its
formulation is a little more simple and deterministic, because we may safely neglect
inaccurately quantified nongravitational torques. At the same time, the analysis
provides us a clear response as to whether the current rotation state of (208) Lacrimosa
occupies the Slivan state or not. Equipped with this knowledge, we can then explore
possibilities of very long-term evolutionary scenarios for Lacrimosa (1~Gyr timescale;
Section~\ref{ev2}), although in this case with less determinism.

\subsection{Short-term spin state evolution of (208) Lacrimosa} \label{ev1}
The sidereal rotation frequency $\omega$ is conserved when restricting to the secular
effects of the solar gravitational torque. Consequently, the only evolving component of
the rotation state is the direction ${\bf s}$ of the spin vector. As discussed 
in Appendix~\ref{appa1}, the flow of ${\bf s}$ on a unit celestial sphere may be
understood using the Colombo top model. The orbital precession frequency of interest
may be either the forced frequency $s_6\simeq -26.34$ arcsec~yr$^{-1}$ or the proper frequency
$s\simeq -67.25$ arcsec~yr$^{-1}$. As the flow of ${\bf s}$ in the prograde-rotating mode is 
fundamentally affected by the presence of the resonant zone about the Cassini
state~2 (``Cassini resonance''), it is useful to first determine whether or not this
resonance exists. For a given orbit, such as that of (208) Lacrimosa, and the two
possible orbital precession modes, the answer depends on two parameters (more specifically, on
their product $P\,\Delta$): (i) the sidereal rotation period $P$, and (ii) the dynamical
ellipticity $\Delta$. At the current epoch, $P$ is known very accurately. As discussed 
in the previous section, observations constrain $\Delta$ as well, but with a much smaller
accuracy.

Figure~\ref{f1} shows maximum obliquity extension of the Cassini resonance as
a function of $\Delta$ for two different values of the rotation period: (i) the
present value $P=14.085734$~hr (left), and (ii) a twice that value, $P=28$~hr
(right). The latter may correspond to the situation in the distant future, because we
showed that the YORP effect decreases the rotation rate. In the first case, (i),
the Cassini resonance exists for the precession mode $s_6$ whenever $\Delta >
\Delta_\star \simeq 0.217$. As $\Delta$ increases, the location of the Cassini
resonance moves to larger obliquity values and its extension slightly decreases.
The Cassini resonance related to the proper frequency $s$ does not exist for
any value of $\Delta$. In the case of the longer rotation period $P=28$~hr, (ii), 
the onset of the Cassini resonance associated with the $s_6$ frequency moves to 
$\Delta_\star\simeq 0.115$. This is because for a fixed orbital precession frequency 
$\Delta_\star \propto P^{-1}$. The novel feature consists of bifurcation of the
Cassini resonance associated with the $s$ frequency at $\Delta_\star \simeq 0.305$.
This resonance is wider in the obliquity because the proper orbital
inclination $I_{\rm P}$ is about four times larger than the forced inclination
$I_6$. The $s$-frequency Cassini resonance appears at low obliquity values at
$\Delta_\star$, and is well separated from the $s_6$-frequency Cassini resonance.
For $\Delta > 0.4$, on the other hand, the two resonances approach and eventually overlap.
The resonance overlap occurs at the $\Delta$ value which is inversely proportional
to the rotation period; for instance with $\Delta\simeq 0.35$ the required
rotational period is $P\simeq 32$~hr.

Returning to the present rotational configuration of Lacrimosa (left panel of
Figure~\ref{f1}), we now focus on the red symbols: these are just under 1000
solutions described in Section~\ref{data}, all of which correspond to statistically
acceptable fits to the observations. The $95\%$ confidence level interval of the
obliquity ranges from $19.5^\circ$ to $26.9^\circ$, with the best-fit value of
$22.4^\circ$. The same confidence-level interval of the dynamical ellipticity is
in between $0.30$ and $0.39$, with the best-fit value of $0.35$. There is a slight
correlation between these two values, pushing the obliquity to larger values for
smaller ellipticity values. The main take-away message here is that only two stray
solutions out of $1000$ fall into the range of the obliquity values delimiting the
Cassini resonance of the $s_6$ orbit precession mode; the majority of the 1000 solutions,
including the best-fitting solutions, provide dynamical ellipticity values
away from the resonance criterion. Assuming the rotation pole
direction (i.e., obliquity) is set accurately enough, the necessary value of the
ellipticity $\Delta$ would be about $30$\%--$35$\% smaller than the values
determined from the shape models. It is highly unlikely that the shape models
would be mistaken at this level, or that the internal density would deviate so much from a 
uniform distribution. Instead, we may preliminarily conclude that the spin state
of (208) Lacrimosa is not in the Slivan state.
\begin{figure}[t]
 \begin{center} 
 \includegraphics[width=0.48\textwidth]{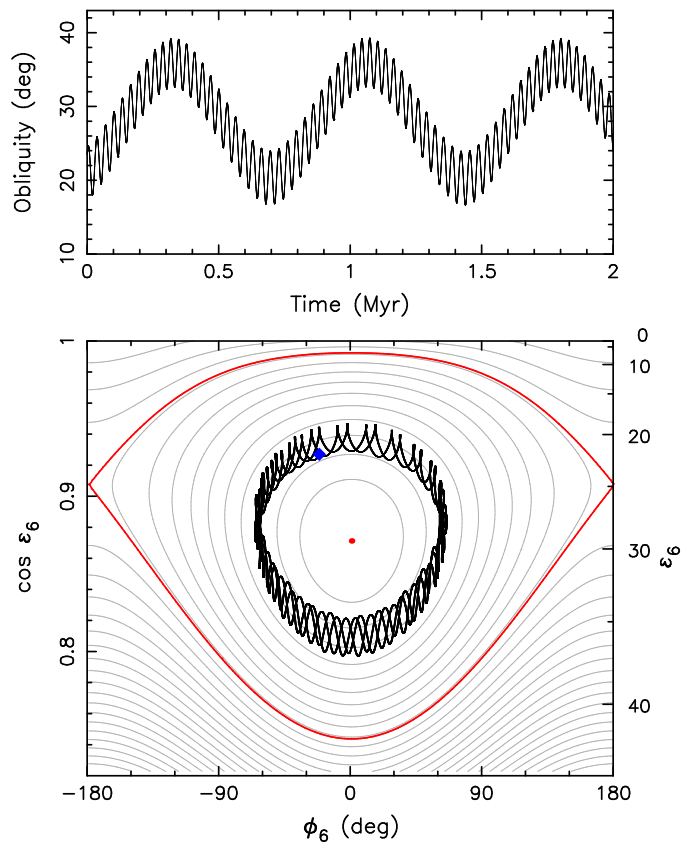}
 \end{center} 
 \caption{\label{f2}
  Top panel: Time evolution of the osculating obliquity $\varepsilon$
  for (208) Lacrimosa over
  the $2$~Myr interval using numerical integration of Eq.~(\ref{sdyn1}) with
  $\vec{T}_{\rm ng}=\vec{0}$. The initial conditions at the present epoch from the
  best-fit rotation state solution ($P=14.085734$~hr, $\lambda=15.2^\circ$ and $b=
  66.9^\circ$) and $\Delta=0.23$. The short-period oscillations are due to the
  proper term of nodal precession with frequency $s$ (which have a period of
  $\simeq 2\pi/(s-s_6)\simeq 32$~kyr). The long-period and large-amplitude
  oscillations of $\simeq 745$~kyr  are due to libration
  about the resonant Cassini state 2 associated with frequency $s_6$ (``the Slivan
  state''). Bottom panel: Phase portrait of the Colombo top model for the
  $s_6$ frequency and precession constant $\alpha\simeq 29.75$ arcsec~yr$^{-1}$ (i.e.,
  $P=14.085734$~hr and $\Delta=0.23$ in Eq.~\ref{alpha}); the ordinate is either
  $\cos\varepsilon_6$ (left) or $\varepsilon_6$ (right) and the abscissa is $\varphi_6$.
  The light-gray curves are isolines of the first integral $C(\varepsilon_6,
  \varphi_6)=$~constant given by Eq.~(\ref{ham}). Critical curves of the spin-orbit
  resonance, namely the separatrix and the stable equilibrium C$_2$, are highlighted in
  red. The black curve is the numerically integrated pole of (208) Lacrimosa from
  the top projected into the plane of these variables; the blue diamond is the
  current position of the pole.}
\end{figure}
\begin{figure}[t]
 \begin{center} 
 \includegraphics[width=0.48\textwidth]{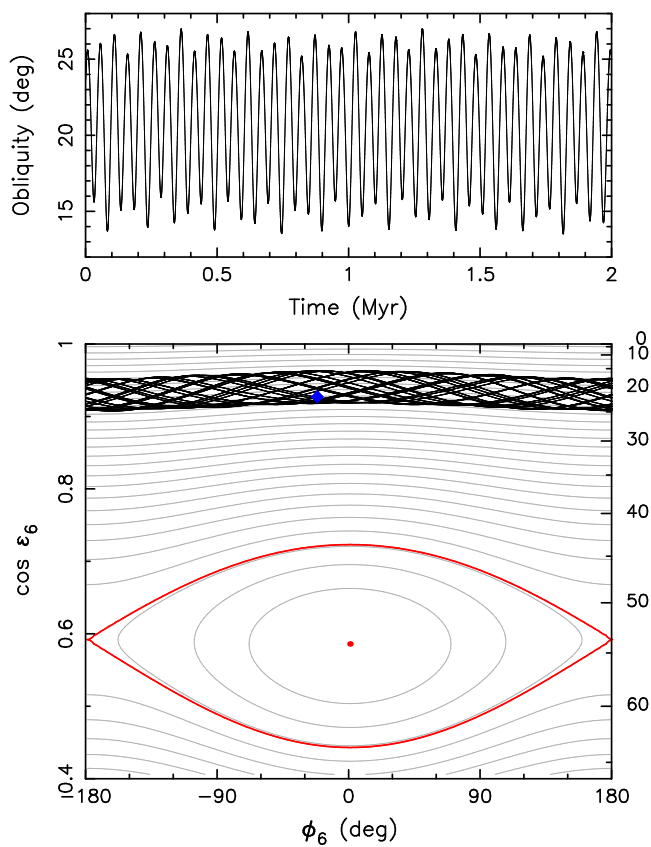}
 \end{center} 
 \caption{\label{f3}
  Top panel: Time evolution of the osculating obliquity $\varepsilon$
  for (208) Lacrimosa over the $2$~Myr interval using numerical integration of
  Eq.~(\ref{sdyn1}) with $\vec{T}_{\rm ng}=\vec{0}$. The initial conditions at the
  present epoch from the best-fit rotation state solution ($P=14.085734$~hr, $\lambda=
  15.2^\circ$ and $b= 66.9^\circ$) and $\Delta=0.35$. The amplitude of the
  oscillations, which is larger than the proper inclination ($\simeq 2.15^\circ$), is forced by the
  Cassini state 2 of the $s$ frequency at $\simeq 6.5^\circ$ (see Figure~\ref{f1}). 
  Bottom panel: Phase portrait of the Colombo top model for the
  $s_6$ frequency and precession constant $\alpha\simeq 45.27$ arcsec~yr$^{-1}$ (i.e.,
  $P=14.085734$~hr and $\Delta=0.35$ in Eq.~\ref{alpha}); the ordinate is either
  $\cos\varepsilon_6$ (left) or $\varepsilon_6$ (right) and the abscissa is $\varphi_6$.
  The light-gray curves are isolines of the first integral $C(\varepsilon_6,
  \varphi_6)=$~constant given by Eq.~(\ref{ham}). Critical curves of the spin-orbit
  resonance, namely the separatrix and the stable equilibrium, are highlighted in
  red. The black curve is the numerically integrated pole of (208) Lacrimosa from
  the top projected into the plane of these variables; the blue diamond is the
  current position of the pole.}
\end{figure}

A more detailed understanding of the situation ---leading to the same conclusion--- is
provided by Figures~\ref{f2} and \ref{f3}. Here we show output from a numerically
integrated spin evolution over the next $2$~Myr. Initial data are from the best-fitting
solution in Section~\ref{data}, namely $(\lambda,\beta)=(15.2^\circ,66.9^\circ)$ and
$P=14.085734$~hr. Results in Figure~\ref{f2} are for dynamical ellipticity $\Delta=0.23$.
This value is incompatible with the shape models fitting the observations, but it is the
value that we predict will match the Slivan-state location. Results in
Figure~\ref{f3} are for the best-fitting dynamical ellipticity $\Delta=0.35$, and confirm
Lacrimosa's spin misalignment with respect to the Slivan state. We used a full-fledged
numerical scheme described in Appendix~\ref{appa2} in which the secular spin evolution is
propagated together with the heliocentric orbital motion. Radiative torques were neglected,
which is an approximation that is well justified by the short interval of time described. 

The upper panels on both Figures~\ref{f2} and \ref{f3} show the osculating obliquity as
a function of time. The bottom panels show the phase space of the Colombo-top model
associated with the $s_6$ precession frequency (see the Appendix~\ref{appa1}): (i) the longitude 
$\varphi_6$ (coordinate) reckoned from the direction $90^\circ$ away from the ascending 
node $\Omega_6=s_6 t + \Omega_{6,0}$ (with $\Omega_{6,0}\simeq 289^\circ$ in the
planetary invariable system and time $t$ origin at J2000.0), and (ii) $\varepsilon_6$
(or $\cos\varepsilon_6$ on the left ordinate; momentum) which is the obliquity value in the
orbital frame with node $\Omega_6$ and inclination $I_6\simeq 0.53^\circ$. The solid black
line in all panels is the result from our numerical propagation. The gray lines in the
bottom panels are isolines of the Colombo model first integral (\ref{ham}). Because 
there are more terms contributing to the precession of Lacrimosa's node, in particular
the proper $s$ term, the gray lines serve only as guidelines of the true motion about
which the solution oscillates. Two particularly interesting isolines of the first
integral are highlighted in red: (i) the separatrix (boundary) of the Cassini resonance,
and (ii) the Cassini state 2 (red dot in the center of the resonant zone).
The spin evolution described in Figure~\ref{f2} confirms what is suggested by Figure~\ref{f1},
namely that a smaller dynamical ellipticity value $\Delta=0.23$ would help to locate the 
spin evolution to the Slivan state. The phase space trajectory librates about the
Cassini state 2. The usefulness of representing the secular spin evolution in this
coordinate system stems from the fact that the Slivan state dictates the principal
features of the motion. In particular, the large-amplitude and long-period
oscillation of the obliquity directly reflects libration motion about the resonance
center C$_2$. The effects related to the leading term in the orbital plane precession, namely
the proper term with frequency $s$, represent only a small perturbation. This is because
the libration period of $\simeq 745$~kyr is an order of magnitude longer than any of the
periods of significant terms characterizing the precessional motion of the orbital plane
in space.
\begin{figure}[t]
 \begin{center} 
 \includegraphics[width=0.48\textwidth]{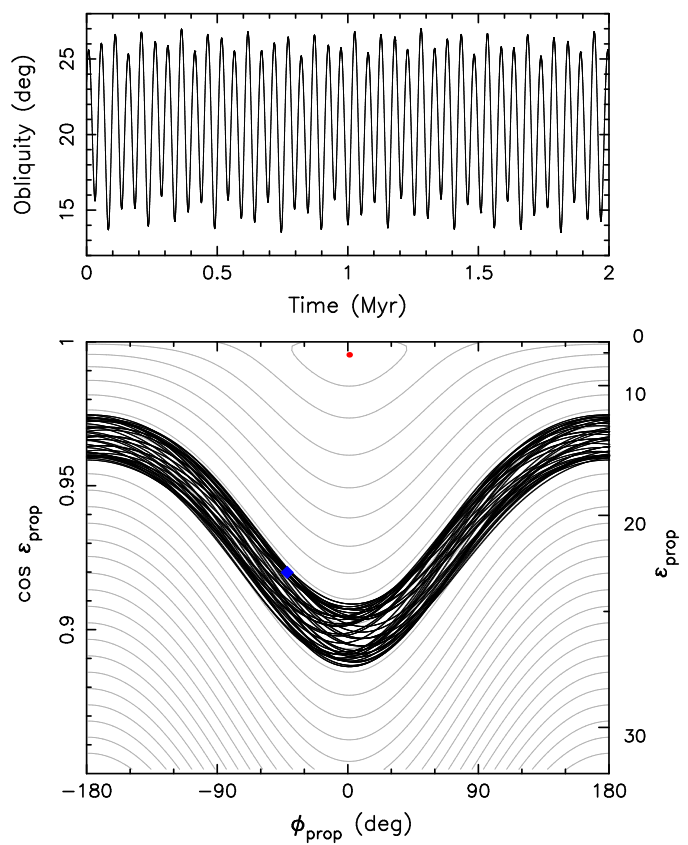}
 \end{center}
 \caption{\label{f4}
  Same as in Figure~\ref{f3}, but here the bottom panel shows phase
  space coordinates of the Colombo top model for the $s$ frequency. As shown in
  Figure~\ref{f1}, the Cassini resonance does not exist and the Cassini state
  2 has an obliquity of $\simeq 6.5^\circ$ (red point). Lacrimosa's spin vector
  circulates about C$_2$ (black line) and follows the isolines of the first
  integral (\ref{ham}) more closely than in Figure~\ref{f3}.}
\end{figure}

However, the observations support a different behavior depicted by Figure~\ref{f3}. In this case,
the Cassini resonance is displaced to larger obliquity and the true evolutionary
path of Lacrimosa's spin simply circulates about the Cassini state 1 (phase space
representation in Figure~\ref{f3} is not suitable to show the location of this center,
which maps onto obliquity $\varepsilon_6\simeq 0.73^\circ$ and $\varphi_6=\pm 180^\circ$).
The osculating obliquity of Lacrimosa (top panel) shows a simple oscillatory
behavior with an amplitude of $\simeq 6.5^\circ$. This value is larger than the 
obliquity oscillation related to the motion about the C$_1$ center and is even larger than
the proper inclination $I_{\rm P}\simeq 2.15^\circ$ of Lacrimosa's orbit. In fact, it
is entirely forced by the obliquity of the Cassini state 2 related to the orbital
plane precession mode with proper frequency $s$ (see left panel of Figure~\ref{f1}).

In order to better understand this effect, we also re-mapped the numerically
determined spin evolution of Lacrimosa to the coordinates of the phase space of
the Colombo-top model associated with the $s$ precession frequency. This is shown
in the bottom panel of Figure~\ref{f4}. The numerically integrated trajectory of
Lacrimosa's spin now more closely follows isolines of the Colombo model first integral
(\ref{ham}), which means the spin evolution is more conveniently represented in these
coordinates. The effects due to the $s_6$ precession mode in the orbital plane
evolution produce only a very small perturbation. The Cassini state 2 (red symbol in
Figure~\ref{f4}) has an obliquity of $\simeq 6.5^\circ$ and its presence triggers
the whole amplitude of the obliquity evolution. The period of the osculating obliquity
oscillations, $\simeq 53$~kyr, is just the period of spin vector circulation about the
Cassini state 2.

We conclude this section by observing that the present-day spin state of (208)~Lacrimosa
is not in the Slivan state despite its prograde sense of rotation. In this respect its
behavior differs from that of the other Koronis family asteroids in this size range. How this is
possible, and its implications for the very long-term evolution of the spin state of this
asteroid are investigated in the following section.
\begin{figure}[pt]
 \begin{center} 
 \includegraphics[width=0.48\textwidth]{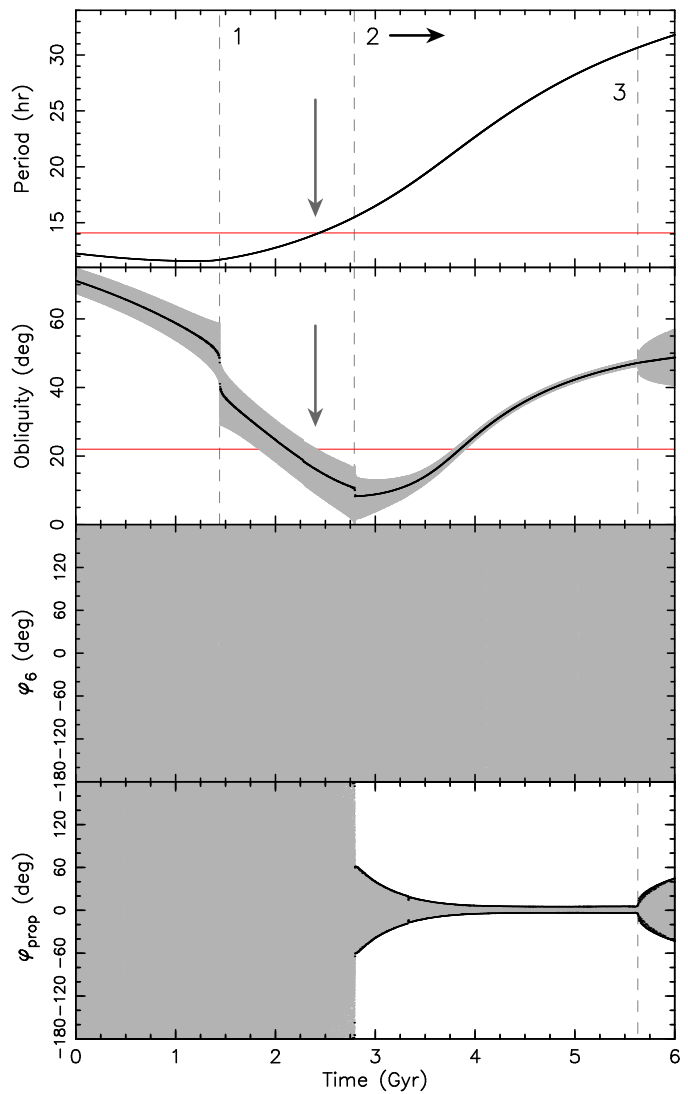}
 \end{center} 
 \caption{\label{f5}
  Example of a possible long-term evolution of the rotation state for
  (208)~Lacrimosa. Rotation period $P$ (top), osculating obliquity $\varepsilon$
  (middle-up), and longitude $\varphi$ in the orbital frame associated with the
  $s_6$-frequency and $s$-frequency term of the nodal precession (middle-down and bottom;
  note $\varphi$ is measured from an axis $90^\circ$ away from the corresponding nodal
  line). The gray dots are densely output osculating values (with a time-step of $5$~kyr).
  Black symbols in the obliquity panel are average values in a $2$~Myr running window;
  black symbols in the bottom panels are maximum and minimum values of the respective
  longitude in a $2$~Myr running window. The dynamical model uses solar gravitational
  torque and the YORP effect with parameters determined from the best-fitting solution
  in Section~\ref{data}. The red lines in the upper two panels show the present state
  of (208)~Lacrimosa for reference. At the epoch of $\simeq 2.4$~Gyr, the propagated spin
  evolution roughly matches the present state (as indicated by the gray arrows). 
  At $\simeq 1.45$~Gyr (vertical dashed line 1), the solution jumps over the
  Slivan state of the $s_6$ precession frequency, where other large Koronis prograde-rotating
  asteroids are located. At $\simeq 2.8$~Gyr (vertical dashed line 2), the solution
  starts to closely follow the Cassini state 2 associated with the $s$ precession frequency.
  This is allowed by (i) the low obliquity (where C$_2$ is located), and (ii) the increasing
  rotation period. The Cassini resonance formally bifurcates when the rotation period
  reaches $\simeq 24.4$~hr, i.e., at $\simeq 4.1$~Gyr. Finally, at $\simeq 5.63$~Gyr
  (vertical dashed line 3), the small-amplitude oscillations about the resonant
  Cassini state 2 in the $s$ precession frequency frame become perturbed by an overlap with the
  Cassini resonance associated with the $s_6$ precession frequency. The simulations had an
  initial rotation period of $12.25$~hr and an initial obliquity of $70^\circ$.}
\end{figure}

\subsection{Possible long-term evolution of the rotation state of (208) Lacrimosa} \label{ev2}
\citet{vetal2003} noted that many but not all initial conditions of possible long-term
evolution scenarios resulted in the Slivan-state situations reported by \cite{slivan02}.
\citet{vetal2003}  showed the positive cases (e.g., Fig.~1 in their paper) but only commented
on the negative cases. For obvious reasons, we are now interested in the opposite situation.

The initial data suitable for capture in the Slivan state had the following common
properties \citep[see][]{vetal2003}: (i) the YORP evolution asymptotically decelerated the
rotation of the asteroid, and (ii) the initial rotation period was smaller than $\simeq 7$~hr.
If these conditions were satisfied, the initial obliquity had only to be positive, but
was not restricted otherwise.
The generic evolution first made the obliquity reach a small value due to the
YORP torque, while still keeping the rotation period short enough. As a result, the
precession frequency $\alpha\cos\varepsilon$ from Eq.~(\ref{precc}) along this evolutionary
path remains smaller than the $-s_6$ frequency. Only when the rotation period increases
sufficently does the resonant condition $\alpha\simeq -s_6$ for small obliquity values become
satisfied (this is because $\alpha\propto P$, Eq.~\ref{precc}).
At the same time, the capture into the resonance is guaranteed (i.e., $100$\% probable)
as long as the resonant condition occurs when the instantaneous obliquity is $\leq
20^\circ$, a comfortably large value. Once captured in the Slivan state, the continuing increase
in the rotation period due to the YORP effect only makes the Cassini resonance drift toward
a larger obliquity, which eventually approaches $\varepsilon\simeq 50^\circ-55^\circ$ where the
YORP-driven period
evolution stalls. We note that the spin state follows this evolution adiabatically, because
the characteristic timescale of the YORP-driven changes is much longer than the libration period
about the Cassini state.
\begin{figure}[t]
 \begin{center} 
 \includegraphics[width=0.48\textwidth]{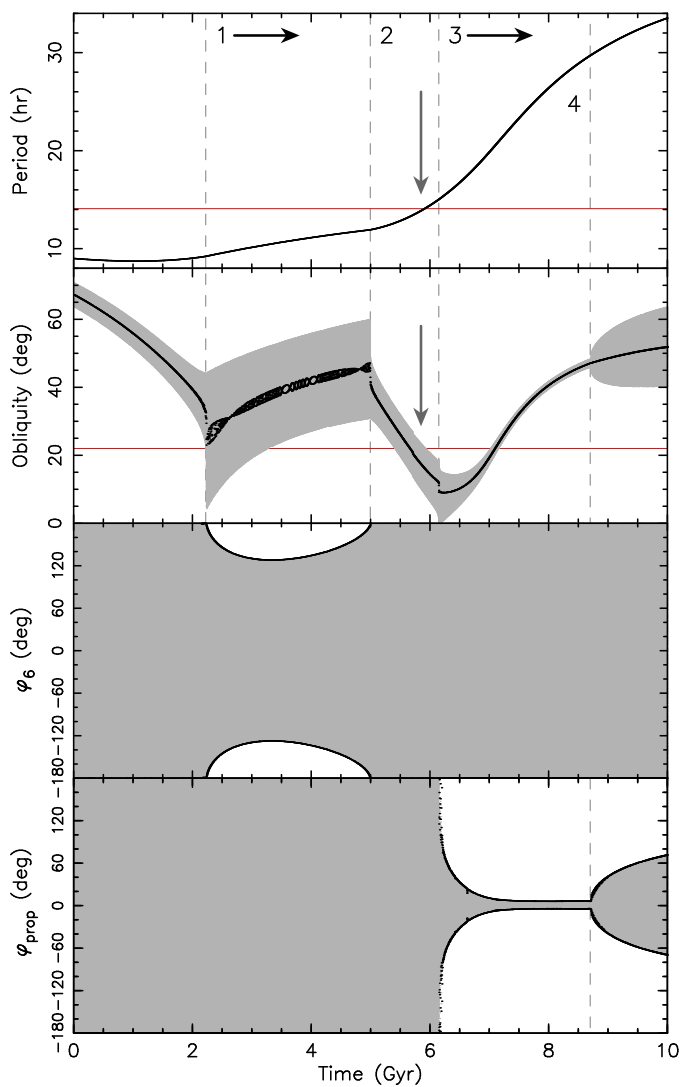}
 \end{center} 
 \caption{\label{f6}
  Same as in Figure~\ref{f5}, but  now for a different initial rotation period
  of $9$~hr. In many respects, the evolution is similar to that shown in the previous
  figure with one important exception: at $\simeq 2.25$~Gyr (vertical dashed line 1) the
  spin state  becomes captured in the Slivan state of the $s_6$ precession frequency; it remains
  located in the Slivan state until $\simeq 5$~Gyr (vertical dashed line 2), when
  the amplitude of resonant libration grows to $180^\circ$. Consequently, the spin state
  leaves the Slivan state and continues to evolve primarily by YORP torques: the
  obliquity drifts to small values and the rotation period slowly increases; at $\simeq
  5.85$~Gyr (highlighted by the arrows), both roughly match the current values of
  (208)~Lacrimosa. As the rotation period continues to grow, the spin evolution follows
  the trend seen also in Figure~\ref{f5}: it starts to closely follow the Cassini state 2
  associated with the $s$ precession frequency (eventually becoming captured in the
  corresponding Slivan state).}
\end{figure}

\begin{figure*}[t]
 \begin{center}
 \includegraphics[width=0.9\textwidth]{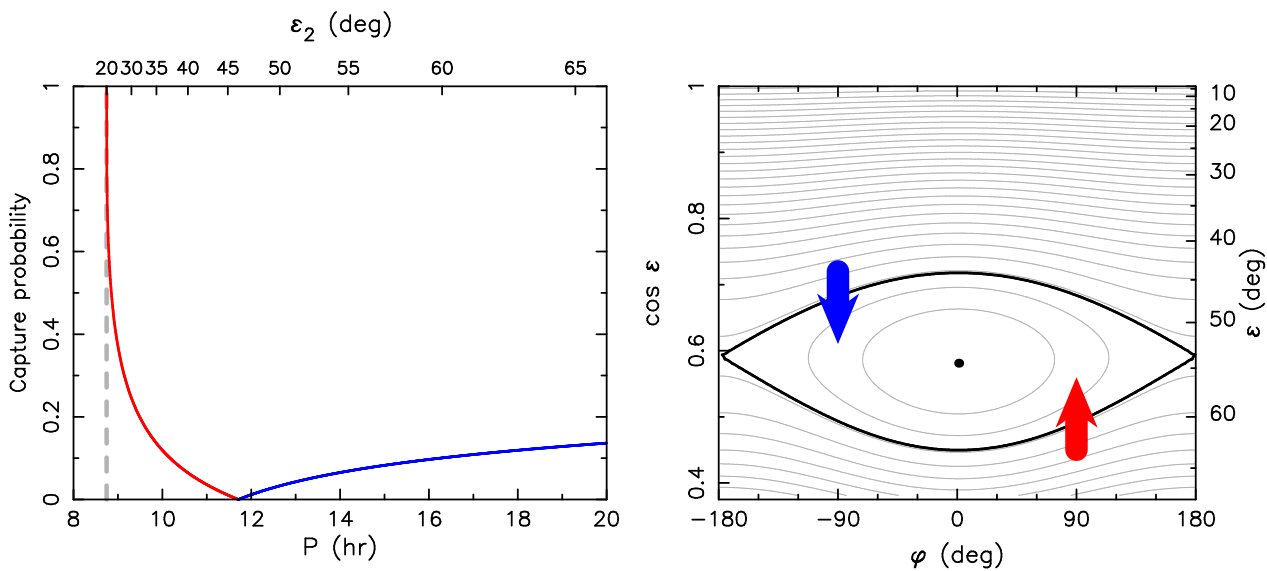}
 \end{center}
 \caption{\label{f7}
  Left panel: Capture probability to the spin-orbit resonance in a Colombo
  top model with an adiabatically slow change in the asteroid rotation period $P$ (at
  the abscissa). Heliocentric orbit of (208) Lacrimosa, dynamical ellipticity $\Delta=
  0.35,$ and $s_6$ mode of the orbital node precession were used. With these assumptions, the
  resonance bifurcates at $\simeq 8.73$~hr
  rotation period (vertical dashed line). The red line indicates probability $P_+$ of 
  a capture from orbits originally circulating about the Cassini state C$_3$, the blue
  line indicates probability $P_-$ of a capture from orbits originally circulating about
  the Cassini state C$_1$. The analytical theory of \citet{hm1987} is briefly recalled in the
  Appendix; see Eqs.~(\ref{prob}) and (\ref{psi}). The upper abscissa shows obliquity
  $\varepsilon_2$ of the Cassini state C$_2$, the equilibrium point of the resonance.
  Right panel: Phase portrait of the Colombo top for rotation period $P=14.085734$~hr
  (other parameters as above). The gray curves are isolines of the integral
  $C(\varepsilon,\varphi)=$~constant given by Eq.~(\ref{ham}). The black curve is the
  separatrix and the black dot shows the location of the Cassini state C$_2$. The arrows schematically
  indicate  the capture in the resonance from orbits originally circulating about
  the Cassini state C$_1$ (blue) and C$_3$ (red); in a model where only the rotation
  period $P$ slowly changes, the former occurs for a decrease in $P$ and the latter occurs for an
  increase in $P$.}
\end{figure*}

What happens in the situation where (i) in the above paragraph is satisfied, but (ii) is not
satisfied (i.e., the initial rotation period is longer)? An example of such evolution is shown in
Fig.~\ref{f5}. In this case, we assume $P=12.25$~hr and $\varepsilon=70^\circ$ initially,
and let the evolution proceed with the YORP torques characteristic of Lacrimosa, i.e.,
body of $D\simeq 44$~km size and $2$ g~cm$^{-3}$ bulk density. We used $\Delta=0.35$, which is
Lacrimosa's nominal value of
dynamical ellipticity (Fig.~\ref{f1}). The initial phase of the evolution resembles what
has been described above: the YORP torque causes the obliquity to decrease, while the rotation
period evolves slower (this is because near $\varepsilon\simeq 55^\circ$ the rotation
period change due
to YORP is nil). However, the main difference is that already the initial value of
the spin axis precession rate $\alpha\cos\varepsilon$ is faster than $-s_6$ because of the
larger $P$ value. At about $1.45$~Gyr, the resonance condition $\alpha\cos\varepsilon
\simeq -s_6$ becomes satisfied. At this moment, the mean obliquity is still large ---about
$50^\circ$--- and the resonance has been approached from the zone of larger obliquity
values (rotation pole circulating about the Cassini state C$_3$). The adiabatic capture
theory described in Appendix~\ref{appa1} \citep[see also][]{h1982} allows us to estimate
the capture probability. Using Eqs.~(\ref{prob}) and (\ref{psi}) we find this probability is
zero (see Fig.~\ref{f7}). Indeed, the numerically propagated spin of  Lacrimosa jumped over
the resonance and continued evolving toward smaller obliquity while  the rotation
period increased due to the YORP torques. At $\simeq 2.4$~Gyr, the possible age of the Koronis family
\citep[e.g.,][]{netal15}, the
simulated obliquity and rotation period closely resemble those of Lacrimosa. In this
view, the lack of Lacrimosa's pole residence in the Slivan state is naturally explained
by avoiding a capture in the first place.

For the sake of interest, we continued our simulation until $6$~Gyr, allowing us to predict
what may possibly happen to Lacrimosa's spin state in the future. At about $2.8$~Gyr,
the simulated spin starts to closely follow the Cassini state C$_2$ associated with the
proper orbital frequency $s$. At small obliquity values, the rotation period continues
to decrease and at about $4.1$~Gyr, when $P\simeq 24.4$~hr, the Cassini resonance of
this frequency bifurcates (see also right panel on Fig.~\ref{f1}, which applies to only
slightly larger rotation period of $28$~hr). From that epoch, the modeled spin state
becomes locked in the new Slivan (resonant) state, but this time associated with the proper
frequency. Because the proper frequency $s$ is larger than $s_6$, the required rotation
period is longer. The evolution follows the pattern known from the theory of classical
Slivan states in \citet{vetal2003}, namely a long-term increase in the obliquity and
rotation period. Finally, at about $5.63$~Gyr the amplitude of obliquity oscillation starts
to increase. This is associated with the increase in the amplitude of oscillation
of the resonant libration angle (bottom panel at Fig.~\ref{f5}). This phase of evolution
is triggered by an overlap of the Cassini resonances associated with the $s_6$ and $s$
orbital frequencies, which were separated until that moment.

An interesting intermediate case of possible long-term spin evolution is shown in
Fig.~\ref{f6}. We kept the same initial conditions, and other parameters, as above
\citep[Fig.~\ref{f5}, except for the shorter initial rotation period of $P=9$~hr to expect
a regular evolution that would result in a capture in the Slivan
state, ][]{vetal2003}. Because of the shorter $P$ value in the initial phase of the
evolution, the resonance condition $\alpha\cos\varepsilon\simeq -s_6$ is now met in
the situation where obliquity $\varepsilon$ has already evolved to a smaller value of $\simeq
20^\circ$. As a consequence (see Fig.~\ref{f7}), the capture in the Slivan state is
possible at $\simeq 2.25$~Gyr. However, the condition is just barely satisfied and
the capture results in a large-amplitude libration situation about the Cassini state C$_2$.
Subsequently, the evolution takes the usual direction towards larger obliquity while
being characterized by the Slivan state capture. However, the large-amplitude libration
state is susceptible to instability, and the spin state is released from the resonance
followed by an interval of time dominated by YORP torques, during which the obliquity
again migrates toward the smaller value. At $\simeq 5.85$~Gyr, the obliquity and rotation
period match those of Lacrimosa. Obviously, this cannot be accepted as a satisfactory
history for this object, because the needed timescale is longer than the age of the Solar
System. However, smaller members in the Koronis family with a similar rotation
state as in (208)~Lacrimosa, such as (263)~Dresda, could take the evolutionary path described
in Fig.~\ref{f6}. This is because, for them, the YORP torques are stronger and the
associated characteristic timescale of evolution scales $\propto D^2$. Therefore, with a size of about
$26$~km, Dresda's spin evolves due to YORP about $2.8$ times faster. The $5.85$~Gyr
then recalibrates to $\simeq 2.1$~Gyr, plausible for the Koronis family age (we note
that this is obviously just a size-scale argument, because the shape of Dresda may lead
to YORP torques of somewhat different strength). While these details are important
for specific cases, they do not invalidate a general conclusion that some smaller members 
(say, $15-25$~km in size) in the Koronis family might have undergone the spin evolution
depicted in Fig.~\ref{f6}. The interesting difference from larger objects consists of
the past capture in the Slivan state, but later evolution away from it. This is  especially expected
to happen among the smaller Koronis members for which the YORP torques are
stronger. The take-away message is that as the spin states of the smaller members of the Koronis
family become known in the future, we may expect more cases unrelated to
the Slivan pattern seen in the population of the larger Koronis objects 
\citep{slivan02}. It is interesting to note that the spin evolution shown in Fig.~\ref{f6}
evolves also to the capture in the Slivan state associated with the $s$ rather than
$s_6$ frequency. As a result, we may also expect the future spin-state solutions for small
Koronis  members to bring evidence of this configuration.

For sake of completeness we mention that numerical tests with initial rotation period
larger than $16$~hr did not lead to configurations that would match Lacrimosa's
rotation parameters in $2-4$~Gyr.

\section{Conclusions} \label{concl}
In this paper, we present new observations of asteroid (208)~Lacrimosa, one of the
largest members of the Koronis family. When joining these new data with
the previously available photometric dataset, we confirm (and improve) the rotation state
solution obtained earlier by \citet{detal2019}. Unlike in \citet{slivan02}, the rotation
of Lacrimosa is found to be prograde. While \citet{detal2019} still had
two possible pole solutions separated by $180^\circ$ in the ecliptic longitude, here
we find that stellar occultation data allow one of them to be favored. Our
analysis indicates a rotation period of $P=14.085734\pm 0.000007$~hr and a pole direction
in ecliptic longitude and latitude of $(\lambda,\beta)=(15^\circ\pm 2^\circ,67^\circ\pm
2^\circ)$. Thermal and occultation data also effectively constrain Lacrimosa's
volumic size to $D=44\pm 2$~km, in good agreement with a previous solution based
on WISE observations.

Large asteroids in the Koronis family, when in prograde rotation, were found to be
locked in the Slivan state \citep[e.g.,][]{slivan02,vetal2003}. Therefore, we analyzed
Lacrimosa's status with respect to this configuration. We find that Lacrimosa's
spin may well be confined to the Slivan state provided the value of the
dynamical ellipticity $\Delta$ is in the range $\simeq (0.22-0.26)$. However,
our convex shape models obtained from the light-curve inversion analysis result
in larger values, namely $\Delta\geq 0.28$. The bootstrap approach to the observation
fit helps us to constrain $\Delta$ to $0.35\pm 0.05$. Therefore, we conclude that
Lacrimosa's rotation pole does not reside in the Slivan state: in other words, its
dynamical ellipticity is too large or its obliquity too small for the rotation pole
to be in the Slivan state (Fig.~\ref{f1}).

We then sought a reason as to why Lacrimosa is different in this respect from other
large Koronis asteroids rotating in a prograde fashion. The easiest solution we find
consists in the assumption that the initial rotation period of Lacrimosa was slightly longer,
notably in the range of $11$ to $15$~hr. For those values, and initial obliquity larger
than $\simeq 50^\circ$, we find that spin evolution avoids capture in the Slivan
state. Instead, it typically reaches the Cassini resonance condition at a still
too high obliquity value and consequently jumps over the Slivan state. Further
evolution toward a small obliquity value explains the current spin configuration of
Lacrimosa. Our numerical simulations also suggest that Koronis members with
slower rotation are efficiently captured in the Slivan state associated with
the proper mode $s$ of the orbital precession, instead of the forced mode $s_6$
(the classical Slivan state).

One of the main purposes of this paper is also to highlight an expected diversity
of spin states among the small asteroids in the Koronis family. While the
large members of this family generally follow the Slivan-state paradigm
\citep{slivan02,vetal2003}, small members --for which the YORP torques
are stronger-- may evolve further. Their possible past Slivan states
might already have been destabilitized, allowing evolution to longer
rotation periods and small obliquities. If pushed even further, a new
type of Slivan state, namely capture in the Cassini resonance associated
with the $s$ precession frequency of the orbits, is also expected. Some
other evolutionary paths may also entirely avoid capture in the traditional
Slivan state by jumping over the Cassini resonance. In summary, small
Koronis members should exhibit a much larger variety of spin states than
would be expected from the Slivan sample of large members. The forthcoming
data from future large-scale surveys will allow this conclusion to be tested.

\begin{acknowledgements}
 This research was supported by the Czech Science Foundation: the work of
 DV through grant 21-11058S, the work of J\v{D} and JH through grant 20-08218S.
 The work of JH has been also supported by the INTER-EXCELLENCE grant
 LTAUSA18093 from the Czech Ministry of Education, Youth, and Sports.
 TRAPPIST is a project funded by the Belgian F.R.S.-FNRS under grant FRFC
 2.5.594.09.F. TRAPPIST-North is a project funded by the University of Li\`ege,
 in collaboration with the Cadi Ayyad University of Marrakech (Morocco).
 E.~Jehin is a FNRS Senior Research Associate.   
\end{acknowledgements}

\begin{appendix}

\section{Methods and numerical tools}
In this Appendix we provide a brief overview of the mathematical formulation and numerical tools
needed for description of an asteroid's rotation state over long periods of time.
This has become a classical chapter of celestial mechanics, and so we mostly refer
to previous publications, where more detailed calculations were performed.

\subsection{Theory} \label{appa1}
Rotational angular momentum $\vec{L}$ of an asteroid evolves as a response
to external torques of both gravitational $\vec{T}_{\rm g}$ and nongravitational
$\vec{T}_{\rm ng}$ origin. Aiming to describe long-term evolution of $\vec{L}$,
we assume $\vec{T}_{\rm g}$ and $\vec{T}_{\rm ng}$ are averaged over rotation
and orbital timescales. For sake of simplicity, we also assume the asteroid
rotates about the shortest axis of the inertia tensor (appropriate for cases
discussed in this paper), therefore $\vec{L}=C\omega\,\vec{s}$ with $C$ the
largest principal value of the inertia tensor, $\omega$ the rotation frequency,
and $\vec{s}$ the unit vector specifying direction of $\vec{L}$. The
gravitational part $\vec{T}_{\rm g}$ is dominated by the effect of the Sun,
in particular quadrupole representation of its tidal field at the location
of the asteroid (higher-multipole contributions and those from planets may be
safely neglected). The nongravitational part $\vec{T}_{\rm ng}$ is due to the
YORP effect. In this model, the gravitational torque may be expressed using a
simple analytical formula \citep[e.g.,][]{bfv2003}. Analytic approaches
for the YORP torques are also available \citep[e.g.,][]{nv2007,nv2008,bm2008},
but they are not practical for our purposes. Rather, we use an averaged representation
of a numerical work presented in \citet{cv2004}.
With all these assumptions adopted, the Euler equation describing secular evolution
of $\vec{L}$ reads
\begin{equation}
 \frac{d\vec{L}}{dt} = -\left[\alpha \left(\vec{c}\cdot\vec{s}\right)
  \vec{c}+\vec{h}\right]\times \vec{L}+\vec{T}_{\rm ng} \; , \label{sdyn1}
\end{equation}
with the first term on the right-hand side  being essentially the gravitational
torque. 

Let us first briefly focus on the effects due to the gravitational torque
(hence, for a moment assuming $\vec{T}_{\rm ng}=\vec{0}$). 
Referring $\vec{L}$ to the inertial space would imply $\vec{h}=\vec{0}$
and $\vec{c}^{\rm T}=\left(\sin I\sin\Omega,-\sin I\cos\Omega,\cos I\right)$,
where $I$ and $\Omega$ are inclination and longitude of the node of the asteroid's
heliocentric orbit. The precession constant $\alpha$ reads
\begin{equation}
 \alpha = \frac{3}{2\eta^3}\frac{n^2}{\omega}\,\Delta, \label{precc}
\end{equation}
where $\eta=\sqrt{1 - e^2}$, $e$ is the orbital eccentricity, $n$ is the orbital
mean motion, and $\Delta$ is dynamical ellipticity of the body defined as
\begin{equation}
 \Delta = \frac{C - \frac{1}{2}\left(A + B\right)}{C}\; . \label{flat}
\end{equation}
Here $(A,B,C)$ ($A\leq B\leq C$) are the principal values of the inertia
tensor. It is useful to note that for the low-eccentricity orbits in the
Koronis family ($a\simeq 2.89$~au and $e\simeq 0.05$) we have \citep[e.g.,][]
{vetal2006c}
\begin{equation}
 \alpha \simeq 55.1\,\Delta\,P_6\;\; {\rm arcsec~yr}^{-1}\;, \label{alpha}
\end{equation}
where $P_6=P/6\,{\rm hr}$ is the rotation period $P$ expressed nondimensionally
in units of $6$~hr (characteristic of many asteroids). The value of $\Delta$ is
restricted to the interval $(0,0.5)$, with most typical values between $0.2$ and
$0.4$ for small asteroids \citep[e.g.,][]{vc2002}.

An alternative to the above-described choice is to refer components of
$\vec{L}$ to the axes comoving with the heliocentric orbital frame of the
asteroid \citep[e.g.,][]{bfv2003,betal2005}. In this case, $\vec{c}$ takes a trivial
form, namely $\vec{c}^T=(0,0,1)$, but now $\vec{h}^T = (\mathcal{A},\mathcal{B},
-2\mathcal{C})$, with
\begin{eqnarray}
 {\mathcal A} & \!\!\! = \!\!\!& \cos \Omega\,{\dot I} -\sin I\sin\Omega \,
  {\dot \Omega} , \nonumber \\
 {\mathcal B} &\!\!\! =\!\!\! & \sin \Omega\,{\dot I} +\sin I\cos\Omega \,
  {\dot \Omega} , \label{abc1} \\
 {\mathcal C} &\!\!\! =\!\!\! & \sin^2 I/2\,{\dot \Omega} ,\nonumber
\end{eqnarray}
where overdots mean time derivatives. In fact, this latter term 
$-\vec{h}\times \vec{L}$ in Eq.~(\ref{sdyn1}) is not of gravitational
origin, but purely induced by transformation to the noninertial, comoving
orbital frame.

In either choice, the gravitational torques alone conserve rotation frequency
$\omega$ and change the spin direction $\vec{s}$ only. If the heliocentric orbit
was fixed in the inertial space (i.e., $I$ and $\Omega$ constant, in particular),
$\vec{s}$ would perform a simple precession about $\vec{c}$ with a frequency
$\dot{\psi} = -\alpha\, (\vec{c}\cdot\vec{s})=-\alpha\cos\varepsilon$. This notation
comes from a traditional representation of $\vec{s}$ in the orbital frame using
\begin{equation}
 \vec{s} = \left( \begin{array}{c}
  \sin\varepsilon\sin\psi \cr \sin\varepsilon\cos\psi \cr \cos\varepsilon \cr
  \end{array} \right), \label{svec}
\end{equation}
where $\varepsilon$ is the obliquity and $\psi$ the precession angle
\citep[e.g.,][]{betal2005}. 

However,  things are more complicated in reality. In our context of secular spin
evolution, this is mainly because the heliocentric orbital plane is not fixed
in the inertial space. On the contrary, the planetary perturbations produce its
complicated evolution which is reflected in time dependence of
$I$ and $\Omega$. It is convenient to merge this information into a complex
and nonsingular variable, $\zeta=\sin I/2 \,\exp(\imath\Omega)$. This is
because $\zeta$ may be represented to an acceptable level of approximation
with a finite number of Fourier terms, namely $\zeta(t)=\sum A_k\exp(\imath
\Omega_k)$, each of which has a constant amplitude $A_k$ (i.e., associated 
inclination value $A_k = \sin I_k/2$) and frequency ${\dot \Omega}_k=s_k$
(therefore $\Omega_k=s_k t+\Omega_{k,0}$). A typical spectrum of frequencies
$s_k$ for an asteroid consists of (i) a proper mode, associated
with free initial conditions of the orbital motion and denoted by $s$,
and (ii) forced modes, imprinted from the perturbing planets (additionally,
terms with frequencies given by linear combinations of $s,$ and planetary
frequencies may also contribute). The forced terms are dominated by effects
of giant planets denoted by $s_6$, $s_7$ and $s_8$. Their numerical values
are $s_6\simeq -26.34$ arcsec~yr$^{-1}$, $s_7\simeq -2.99$ arcsec~yr$^{-1}$ and
$s_8\simeq -0.69$ arcsec~yr$^{-1}$
of consecutively decreasing frequency \citep[e.g.,][]{l1988}. As
the $s_6$-related term reflects primarily perturbations by the gas giants,
Jupiter and Saturn, its amplitude $I_6$ is the largest. As an example, in the
case of (208) Lacrimosa we have $I_6\simeq 0.53^\circ$, while the proper term
has $I_{\rm P}\simeq 2.13^\circ$ and $s\simeq -67.25$ arcsec~yr$^{-1}$.
All other terms in Fourier representation of $\zeta$
have amplitudes at least an order of magnitude smaller. In the first approximation,
we may therefore assume representation of $\zeta$ with only two Fourier terms,
namely (i) the proper term, and (ii) the forced term with the $s_6$ frequency.

The core of complexity related to the moving orbital plane arises from the fact that
the above-mentioned precession frequency $\dot{\psi}$ may enter into a resonance
with some of the frequencies $s_k$ in the Fourier representation of $\zeta$. The
nature of this resonance is best explained in a model where $\zeta$ is represented
with only one Fourier term. In our application of asteroids in the Koronis family,
the more realistic situation with two terms in $\zeta$ may be understood at the
zero order as a high-frequency ($s$) perturbation of the single-term model with
the lower-frequency ($s_6$), or vice versa. This works well especially when the 
two frequencies, $s_6$ and $s$, are well separated.

The single-term model for $\zeta$ is very useful because of its
integrability. This model has been extensively studied and it is known as a Colombo
top problem \citep[e.g.,][]{c1966,hm1987,sail2019,hap2020}. Here we provide its 
most important features relevant to our study.

We assume that $\zeta=\sin I/2\,\exp[\imath(st+\phi)]$, namely the orbital plane 
has a constant inclination $I$ and a node precessing with constant frequency
$s$. The most interesting features of the Colombo top derive from occurrence of stationary
solutions. Their number depends on a nondimensional parameter $\kappa=\alpha/(2s)$.
In a simpler situation, when $|\kappa|< \kappa_\star$, there exists two stationary solutions, 
otherwise there are four stationary solutions (astronomical tradition has it
that we call them Cassini states). The threshold value for $\kappa$ reads 
\citep[e.g.,][]{hm1987,hap2020}
\begin{equation}
 \kappa_\star = \frac{1}{2}\left(\sin^{2/3}I+\cos^{2/3}I\right)^{3/2} \; .
  \label{kappacrit}
\end{equation}
For low-inclination cases, $\kappa_\star\simeq \frac{1}{2}$, and the two new
stationary solutions bifurcate when $\alpha\simeq -s$. While stationary with
respect to the (moving) frame with nodal longitude $\Omega=st+\phi$, the
Cassini states obviously regularly precess in the inertial space. Their 
obliquity value is given by solutions of the equation
\begin{equation}
 \kappa \sin 2\varepsilon = -\sin\left(\varepsilon \mp I\right) \label{oblstat}
,\end{equation}
with the upper sign $-$ for $\varphi=0^\circ$ and lower sign $+$ for 
$\varphi=180^\circ$; the definition of the longitude in the moving frame is
$\varphi=-(\psi+\Omega)$ and it reckons from a direction $90^\circ$ away from the
ascending node \citep[interestingly, the values of $\cos\varepsilon$ for the
Cassini state may be obtained analytically as roots of a quartic equation
derived easily from (\ref{oblstat}); see, e.g.,][]{sail2019,hap2020}.
Of particular interest is $\varphi=0^\circ$ stationary point when 
$|\kappa| > \kappa_\star$ which is usually referred to as the Cassini state 2 
(C$_2$). This is because it has a character of a stable resonant state: small
perturbations make obliquity oscillate about $\varepsilon_2$ and longitude
$\varphi$ librate about zero (see lower panel on Fig.~\ref{f2}). The nature of 
the resonance is seen from
(\ref{oblstat}) whose limit for $I\simeq 0$ becomes $\kappa\sin 2\varepsilon
\simeq -\sin\varepsilon$. The obvious solutions $\varepsilon_1\simeq 0^\circ$ and
$\varepsilon_3\simeq 180^\circ$ correspond to the Cassini states 1 and 3 (to be denoted
C$_1$ and C$_3$), while the Cassini states 2 and 4 are at approximately $\alpha\cos
\varepsilon_{2,4}\simeq -s$. The left-hand side is the regular precession of
${\bf s}$ produced by the gravitational torque of the center, while the right-hand
side is the orbital precession rate. Thus the Cassini-state 2 resonance
expresses 1:1 commensurability between the two. Together with the Cassini-state 4
(C$_4$), C$_2$ form stable and unstable equilibria of the spin-orbit resonance.
The maximum width $\Delta \varepsilon$ of the resonant zone associated with the
Cassini state 2 may be determined from \citep[e.g.,][]{hm1987,wh2004,vetal2006c,
sail2019,hap2020}
\begin{equation}
 \sin\frac{\Delta\varepsilon}{2} =\frac{1}{|\kappa|} \sqrt{\frac{\sin 2I}{\sin
  2\varepsilon_4}}\;, \label{oblw}
\end{equation}
where $\varepsilon_4$ is the obliquity of the unstable equilibrium from 
Eq.~(\ref{oblstat}). Alternatively, one can also use somewhat simpler
\begin{equation}
  \tan\frac{\Delta\varepsilon}{4} =\sqrt{\frac{\tan I}{\tan\varepsilon_4}}\;.
  \label{oblw1}
\end{equation}
An important implication of the square-root factor on the right-hand side of 
(\ref{oblw}) or (\ref{oblw1}) is that $\Delta \varepsilon$ may be
significant (e.g., tens of degrees) even for very small values of $I$ (e.g.,
a degree); see Fig.~\ref{f1} for specific examples.
Another useful aspect of integrability of the Colombo top problem is
the existence of the first integral of motion,
\begin{equation}
 C\left(\varepsilon,\varphi\right)=\kappa \cos^2\varepsilon+\cos I \cos\varepsilon+
  \sin I\sin\varepsilon\cos\varphi \;. \label{ham}
\end{equation}
Conservation of $C(\varepsilon,\varphi)$ allows us to easily represent solutions
in the obliquity ($\varepsilon$) versus longitude $(\varphi$) plane such as those shown
on Fig.~\ref{f2}. Critical points of the surface $C(\varepsilon,\varphi)=$~constant
are obviously the above-mentioned stationary points; in the more interesting case
of a set of four: (i) the minima specify location of C$_1$ and C$_3$, (ii) C$_2$ is
the maximum, and (iii) C$_4$ is the saddle point.

When some of the parameters of the Colombo top model vary slowly in time, 
$C(\varepsilon,\varphi)$ is not strictly constant. Rather, the system slowly drifts
among solutions approximately conserving this parameter. A special situation happens
when the motion approaches the separatrix of the spin-orbit resonance. At this moment,
the future evolution may either (i) avoid the resonance and continue to circulate about
either C$_1$ or C$_3$ equilibrium states, or (ii) it may be captured in the resonance
(thus librating about the Cassini state C$_2$). The process is inherently chaotic 
(unpredictable). Nevertheless, in an adiabatic model it can be approached 
probabilistically \citep[e.g.,][]{h1982}. Surprisingly, all necessary algebra may be 
carried out analytically in the Colombo top model \citep[e.g.,][]{hm1987}.
Assume, as an example, the rotation period of an asteroid slowly changes. This
is reflected in a slow change of the precession constant $\alpha$ in Eq.~(\ref{precc}).
Following the elegant formulation in \citet{hm1987}, one can determine resonance capture
probability $P_+$ of a transition from the solution circulating about C$_3$ (see
the sense of the red arrow in Fig.~\ref{f7}) and resonance capture probability $P_-$ of a 
transition from the solution circulating about C$_1$ (see the sense of the blue arrow 
in Fig.~\ref{f7}). In fact, both may be given using a compact formulation:
\begin{equation}
 P_\pm = {\rm max} \left(\frac{\Psi}{\pm 1 +\frac{1}{2}\Psi},0\right)\; , \label{prob}
\end{equation}
where
\begin{eqnarray}
 \Psi & = & \frac{4}{\pi}\, \Biggl\{\arcsin \left[\frac{\tan (\Delta\varepsilon/4)}
  {\tan\varepsilon_4}\right] - \label{psi} \\
  & & \quad\, 2\, \frac{|\kappa|\sin (\Delta\varepsilon/4)}{\cos I} \sqrt{\sin^2 \varepsilon_4
  - \sin^2(\Delta\varepsilon/4)}\Biggr\} \; .  \nonumber 
\end{eqnarray}
In our context, the rotation period of an asteroid is changed by the YORP effect
(Eq.~\ref{sdyn2}). However, the above-given results are only approximate. This is because
the YORP effect directly changes also the obliquity (Eq.~\ref{sdyn3}), namely one
of the active variables in the Colombo top model. Hence, results from numerical
simulations are needed to verify the capture probabilities given above.

The symplectic numerical scheme of \citet{betal2005} allows, aside from quadrupole
solar torque, to include a general weak dissipative torque $\vec{T}$. In our
case, $\vec{T}=\vec{T}_{\rm ng}$ represents the YORP effect. A distinctive feature 
of the YORP
effect is its ability to change the rotation rate of the asteroid in the long term.
This is associated with the nonzero along-spin component of the torque, namely
(see Eq.~\ref{sdyn1}),
\begin{equation}
 \left(\frac{d\omega}{dt}\right)_{\rm ng} = \frac{\vec{T}_{\rm ng}\cdot \vec{s}}{C}
  \; .\label{sdyn2}
\end{equation}
The YORP effect also acts on $\vec{s}$, in particular obliquity $\varepsilon$ and
precession angle $\psi$ \citep[e.g., Eqs. 5-8 in][]{cv2004}. However, the latter 
represents only a small perturbation compared to the effect produced by the 
gravitational torque. Therefore, we neglect this component and include the YORP effect 
on obliquity only:
\begin{equation}
 \left(\frac{d\cos\varepsilon}{dt}\right)_{\rm ng} = \frac{\vec{T}_{\rm ng}\cdot
  \vec{c}}{C\omega} - \frac{\cos\varepsilon}{\omega}\left(\frac{d\omega}{dt}
  \right)_{\rm ng} \label{sdyn3}
;\end{equation}
(we note that this is conveniently the third component of $\vec{s}$ in our representation
by Eq.~\ref{svec}). Because in this work we aim to illustrate the likely processes
in the Koronis family, we do not need highly accurate determination of the YORP
effect. We consider the YORP strength determined for the nominal (best-fit) model
of Lacrimosa from Section~\ref{data}. Instead of computing YORP torque for
a spin-orbit configuration at a given moment during the numerical simulation, we
follow the approach of \citet{vc2002} and \citet{cv2004}. In particular, we
pre-computed values of the factors $(\vec{T}_{\rm ng}\cdot \vec{s})/C$ and
$(\vec{T}_{\rm ng}\cdot \vec{c})/C -\cos\varepsilon\, (d\omega/dt)_{\rm ng}$ on the 
right-hand side of Eqs.~(\ref{sdyn2}) and (\ref{sdyn3}) as a function of obliquity 
$\varepsilon$ (we note the basic YORP theory does not assume them to be a function 
of $\omega$). We used a sufficiently dense grid of two degrees in obliquity
(see Section~\ref{yorp}).
When performing our long-term spin simulations we simply interpolated these
rotation-rate and obliquity YORP torques.

\subsection{Numerical implementation} \label{appa2}
We implemented the algorithm developed in \citet{betal2005} to numerically
integrate Eq.~(\ref{sdyn1}) (in particular, we use their LP2 splitting
scheme). In our approach, the components of $\vec{L}$ are represented
with respect to the frame comoving with the heliocentric orbit. As
we deal with secular evolution of $\vec{L}$, we may use a  long-enough
time-step of $50$~yr. In addition to initial conditions and dynamical ellipticity
$\Delta$ as the only external parameter, the code needs information
about the orbital evolution due to planetary perturbations. To that end
we use two methods.

In the first, more detailed method used in Section~\ref{ev1}, we determine 
osculating orbital
parameters, in particular semimajor axis $a$, eccentricity $e$, inclination
$I,$ and longitude of node $\Omega$ (all needed in Eq.~\ref{sdyn1}), using
direct numerical integration of the asteroid's heliocentric motion. For that
purpose we adapted the widely known and well-tested integration package%
\footnote{\swift}
{\tt swift}. Because {\tt swift} integrates the full system of equations
of motion for both planets and asteroid(s) it requires an accordingly
short time-step. We used 3~days, short enough to realistically
describe orbital evolution of all bodies (including planet Mercury).
Initial orbital state vectors for the chosen asteroids and a given epoch
were taken from the {\tt AstDyS} internet database,%
\footnote{\httppisa}
and for the planets from the JPL DE405 ephemerides file. To organize the
propagation efficiently, we embedded our secular spin integration
scheme into the {\tt swift} package. This arrangement not only allows to propagate
the spin evolution online, avoiding large output files with the orbital
evolution, but also allows to simultaneously propagate  the spin evolution
of more asteroids or parametric variants of the same asteroid (for
instance testing evolution for different values of the dynamical
ellipticity parameter $\Delta$). We note that the spin propagation only
needs at a given time to know the orbital parameters in the neighboring
grid points in time, which are readily provided by the {\tt swift} integrator.

The above-mentioned implementation is very precise and has been used for
short-term tests such as those shown on Figs.~\ref{f2} -- \ref{f4}. However, it is unnecessarily detailed for the propose of very long-term simulations,
where our goal is to demonstrate the possible evolutionary tracks of Lacrimosa's
spin state over very long timescales (Section~\ref{ev2}, e.g., Figs.~\ref{f5} and
\ref{f6}). This is
because the
implementation based on {\tt swift} code requires a rather short time-step of the
order of days. Therefore, to fully profit from a possibility of a longer time-step
(order of years or so) for the propagation of $\vec{L}$, we also adopted
an approximate variant where the heliocentric orbit evolution was simplified.
This means the semimajor axis and eccentricity were assumed constant (and
equal to the proper elements of the asteroid), and $\zeta$ was represented
with two Fourier terms, namely the proper term and the $s_6$-frequency term
(as discussed above). In this case we also adopted our simplified approach to
the YORP effect, namely interpolating the rotation-rate and obliquity torques
precomputed using the shape model of Lacrimosa (see Section~\ref{yorp}). For
this task, we wrote our own numerical code that implements spin propagator
described in \citet{betal2005}.

\section{Observations using TRAPPIST system}
TRAPPIST-North (TN) and -South (TS) are $0.6$-m Ritchey-Chr\'{e}tien robotic 
telescopes operating at f/8 on German equatorial mounts \citep{jeh2011}.
TN is located at the Oukaimeden Observatory in Morocco (Z53) and the camera 
is an Andor IKONL BEX2 DD ($0.60^{\prime\prime}$/pixel, $20^\prime\times 20^\prime$
field of view). TS is located at the La Silla Observatory in Chile (I40) and the 
camera is a FLI ProLine 3041-BB ($0.64^{\prime\prime}$/pixel, $22^\prime\times
22^\prime$ field of view). 
We observed Lacrimosa in 2020 using the Johnson-Cousins Rc filter in March and 
June and obtained dense lightcurves at solar phase angles of $\sim 19^\circ$ and
$\sim 7^\circ$ respectively (Table~\ref{tab:aspect}).
The images were first calibrated with IRAF scripts using the corresponding 
flat fields, bias, and dark frames. The differential photometry was then performed 
using Python scripts by selecting nonvariable comparison stars with high S/N and 
by testing various aperture sizes.

\section{Model fit to the observations}
\begin{figure*}[t]
 \begin{center}
 \includegraphics[width=0.9\textwidth]{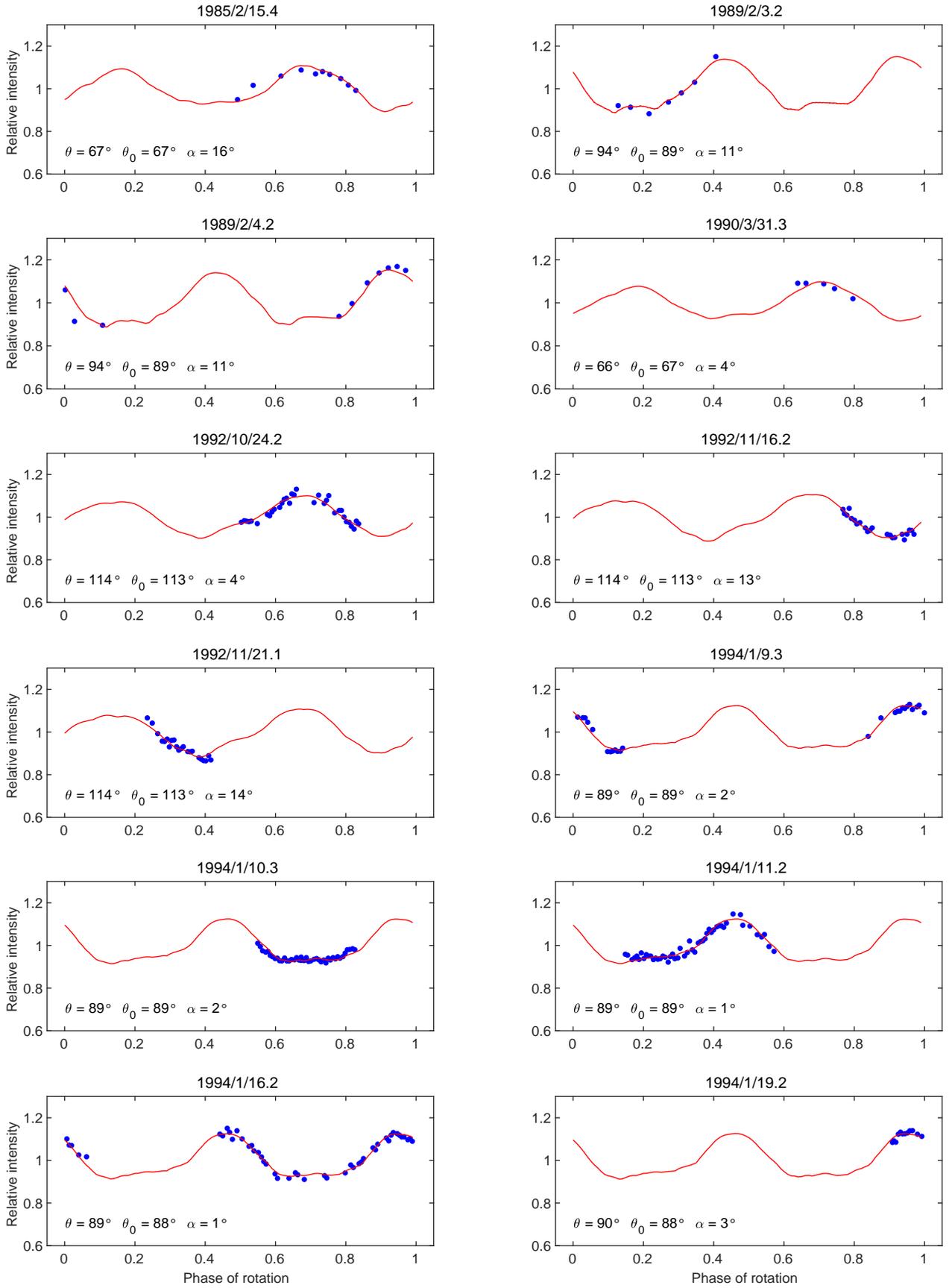}
 \end{center}
 \caption{\label{lc1}
  Observed light curves of Lacrimosa (blue points) shown with synthetic 
  light curves corresponding to the best-fitting model with the pole direction 
  $(15^\circ, 67^\circ)$ and rotation period $14.085734$~hr (red curves). The
  viewing and illumination geometry is described by the aspect angle $\theta$,
  the solar aspect angle $\theta_0$, and the solar phase angle $\alpha$.}
\end{figure*}

\begin{figure*}[t]
 \begin{center}
 \includegraphics[width=0.9\textwidth]{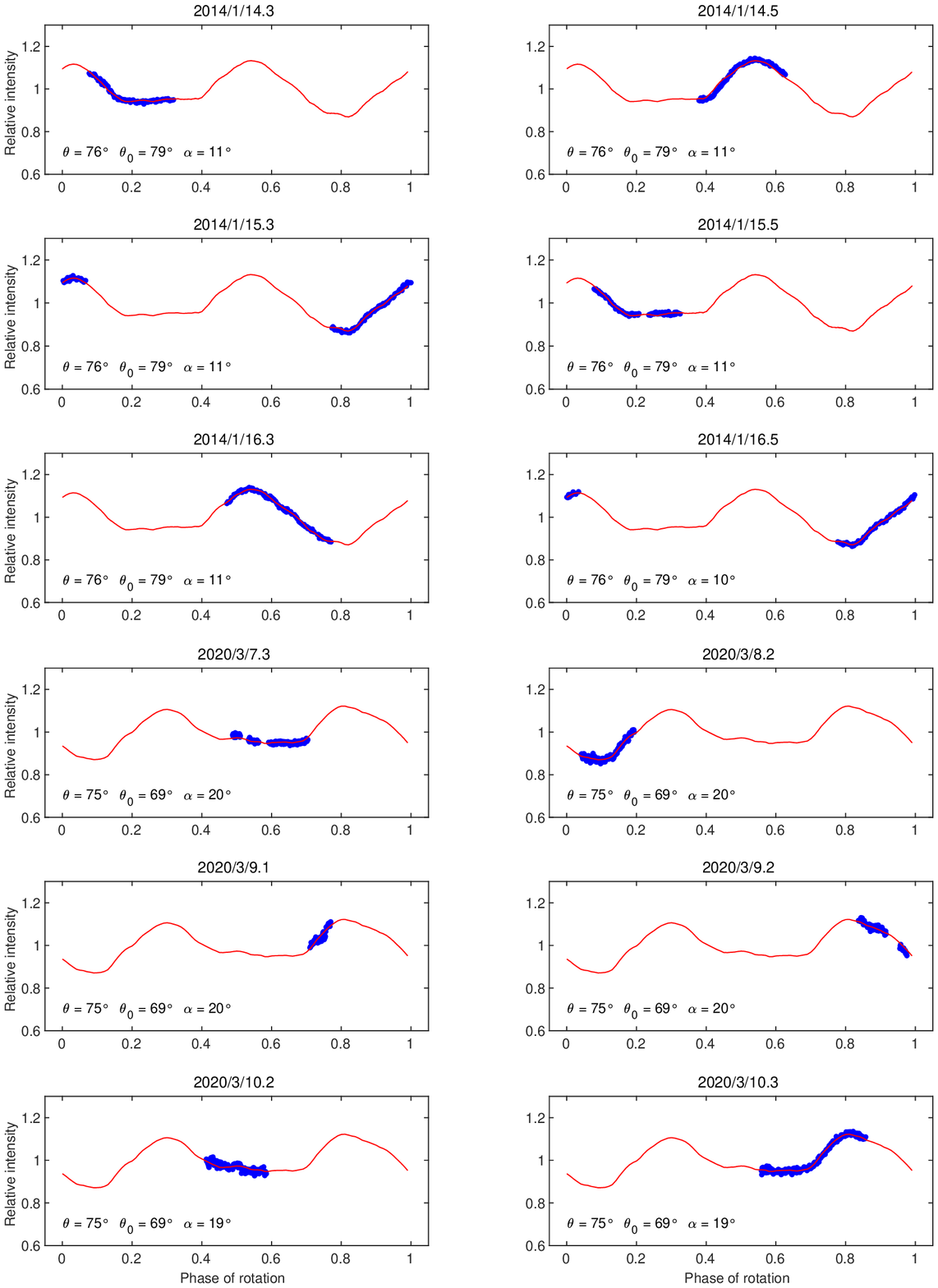}
 \end{center}
 \caption{\label{lc2}
  Observed light curves of Lacrimosa (blue points) shown with synthetic 
  light curves corresponding to the best-fitting model with the pole direction 
  $(15^\circ, 67^\circ)$ and rotation period $14.085734$~hr (red curves). The
  viewing and illumination geometry is described by the aspect angle $\theta$,
  the solar aspect angle $\theta_0$, and the solar phase angle $\alpha$.}
\end{figure*}

\begin{figure*}[t]
 \begin{center}
 \includegraphics[width=0.9\textwidth]{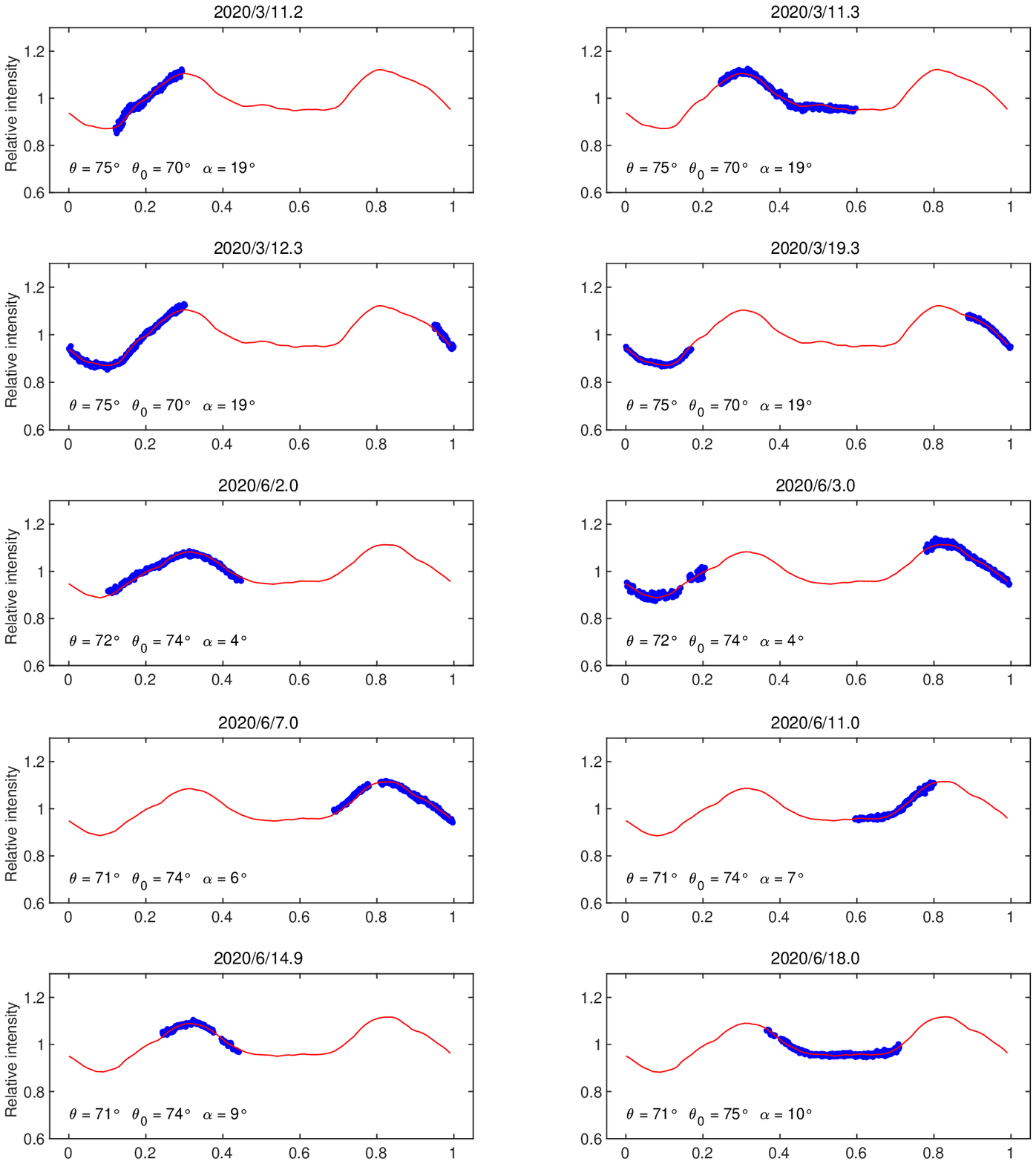}
 \end{center}
 \caption{\label{lc3}
  Observed light curves of Lacrimosa (blue points) shown with synthetic 
  light curves corresponding to the best-fitting model with the pole direction 
  $(15^\circ, 67^\circ)$ and rotation period $14.085734$~hr (red curves). The
  viewing and illumination geometry is described by the aspect angle $\theta$,
  the solar aspect angle $\theta_0$, and the solar phase angle $\alpha$.}
\end{figure*}

\begin{figure*}[t]
 \begin{center}
 \includegraphics[width=0.9\textwidth]{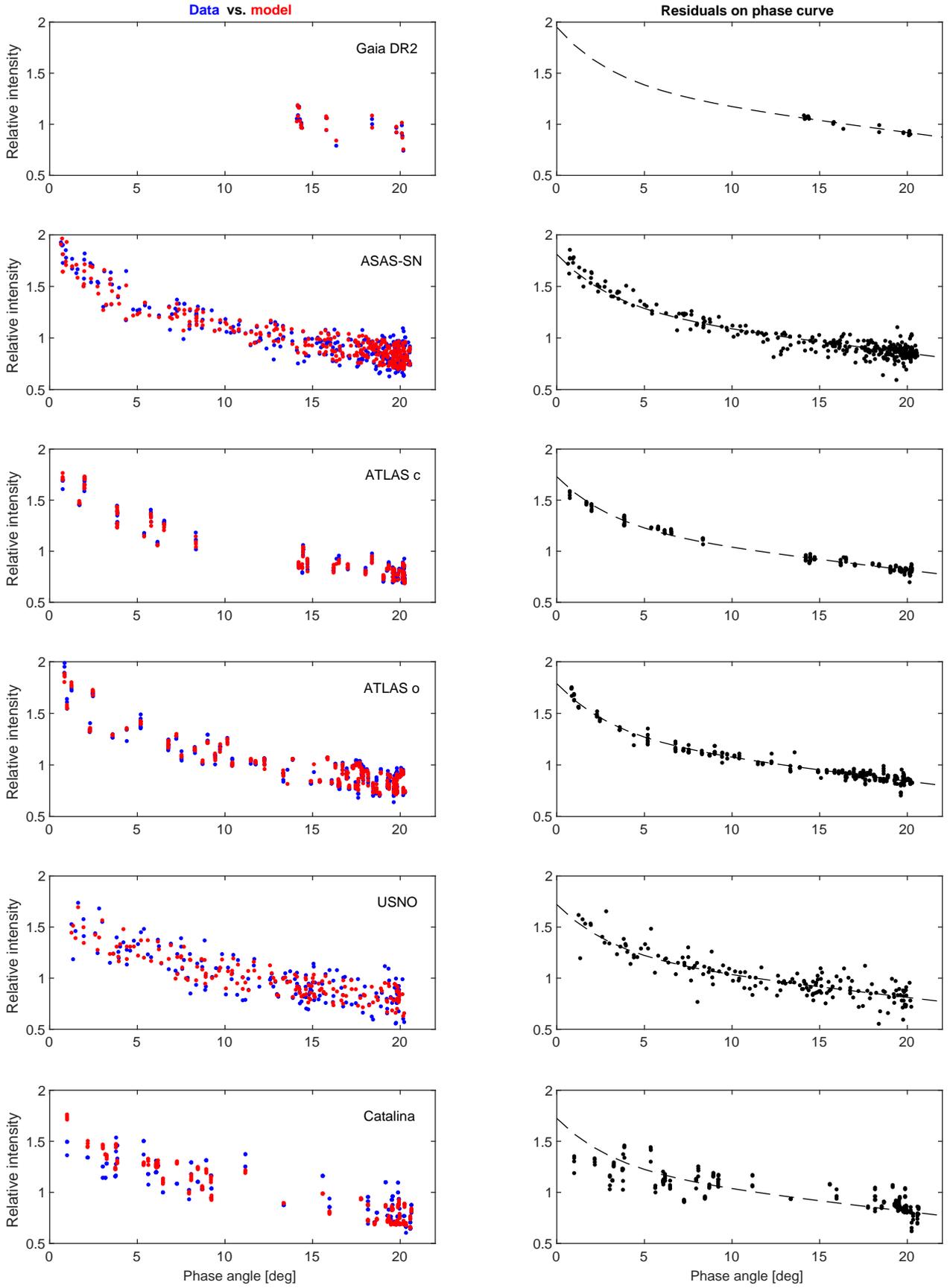}
 \end{center}
 \caption{\label{lcs}
  Left: Observed sparse photometric data with their brightness reduced
  to a unit distance from the Sun and the Earth (blue points) and synthetic data
  produced by the best-fitting model with the pole direction $(15^\circ, 67^\circ)$ and
  rotation period $14.085734$~hr (red points). Right: Residuals (the
  difference between data and model) plotted on the model phase curve (dashed curve).}
\end{figure*}

In this Appendix we show performance of the model using the best-fitting parameters
versus observations listed in Table~\ref{tab:aspect}. Figures~\ref{lc1} to \ref{lc3}
show the traditional light curves, i.e., dense photometry data. We note that the
rotation state solution in \citet{slivan02} and \citet{setal2003} was based on
observations shown in Fig.~\ref{lc1} only. The relative brightness on the
vertical axis is scaled to have the mean value of one. The red curve is the prediction
from our model, the blue symbols are observations. All data are treated as relative
photometry. Figure~\ref{lcs} shows sparse
photometry data from various surveys: blue symbols are the individual observations,
red symbols are the model predictions. The right panels show residuals and the
phase curve (dashed line). 

\end{appendix}


\begin{thebibliography}{}
\expandafter\ifx\csname natexlab\endcsname\relax\def\natexlab#1{#1}\fi

\bibitem[{{Bertotti} {et~al.}(2003){Bertotti}, {Farinella}, \&
  {Vokrouhlick{\'y}}}]{bfv2003}
{Bertotti}, B., {Farinella}, P., \& {Vokrouhlick{\'y}}, D. 2003, {Physics of
  the Solar System - Dynamics and Evolution, Space Physics, and Spacetime
  Structure.} (Kluwer Academic Press, Dordrecht)

\bibitem[{{Binzel}(1987)}]{bin1987}
{Binzel}, R.~P. 1987, \icarus, 72, 135

\bibitem[{{Breiter} \& {Michalska}(2008)}]{bm2008}
{Breiter}, S. \& {Michalska}, H. 2008, \mnras, 388, 927

\bibitem[{{Breiter} {et~al.}(2005){Breiter}, {Nesvorn{\'y}}, \&
  {Vokrouhlick{\'y}}}]{betal2005}
{Breiter}, S., {Nesvorn{\'y}}, D., \& {Vokrouhlick{\'y}}, D. 2005, \aj, 130,
  1267

\bibitem[{{Colombo}(1966)}]{c1966}
{Colombo}, G. 1966, \aj, 71, 891

\bibitem[{{{\v C}apek} \& {Vokrouhlick{\'y}}(2004)}]{cv2004}
{{\v C}apek}, D. \& {Vokrouhlick{\'y}}, D. 2004, \icarus, 172, 526

\bibitem[{{\v{D}urech} {et~al.}(2015){\v{D}urech}, {Carry}, {Delb\`o},
  {Kaasalainen}, \& {Viikinkoski}}]{detal15}
{\v{D}urech}, J., {Carry}, B., {Delb\`o}, M., {Kaasalainen}, M., \&
  {Viikinkoski}, M. 2015, in Asteroids IV, ed. P.~{Michel}, F.~E. {DeMeo}, \&
  W.~F. {Bottke}, 183--202

\bibitem[{{{\v{D}}urech} {et~al.}(2017){{\v{D}}urech}, {Delbo'}, {Carry},
  {Hanu{\v{s}}}, \& {Al{\'\i}-Lagoa}}]{detal2017}
{{\v{D}}urech}, J., {Delbo'}, M., {Carry}, B., {Hanu{\v{s}}}, J., \&
  {Al{\'\i}-Lagoa}, V. 2017, \aap, 604, A27

\bibitem[{{{\v{D}}urech} {et~al.}(2019){{\v{D}}urech}, {Hanu{\v{s}}}, \&
  {Van{\v{c}}o}}]{detal2019}
{{\v{D}}urech}, J., {Hanu{\v{s}}}, J., \& {Van{\v{c}}o}, R. 2019, \aap, 631, A2

\bibitem[{{{\v{D}}urech} {et~al.}(2011){{\v{D}}urech}, {Kaasalainen}, {Herald},
  {Dunham}, {Timerson}, {Hanu{\v{s}}}, {Frappa}, {Talbot}, {Hayamizu},
  {Warner}, {Pilcher}, \& {Gal{\'a}d}}]{detal2011}
{{\v{D}}urech}, J., {Kaasalainen}, M., {Herald}, D., {et~al.} 2011, \icarus,
  214, 652

\bibitem[{{\v{D}urech} {et~al.}(2010){\v{D}urech}, {Sidorin}, \&
  {Kaasalainen}}]{damit}
{\v{D}urech}, J., {Sidorin}, V., \& {Kaasalainen}, M. 2010, \aap, 513, A46

\bibitem[{{Gaia Collaboration} {et~al.}(2018){Gaia Collaboration}, {Spoto},
  {Tanga}, {Mignard}, {Berthier}, {Carry}, {Cellino}, {Dell'Oro}, {Hestroffer},
  {Muinonen}, {Pauwels}, {Petit}, {David}, {De Angeli}, {Delbo},
  {Fr{\'e}zouls}, {Galluccio}, {Granvik}, {Guiraud}, {Hern{\'a}ndez},
  {Ord{\'e}novic}, {Portell}, {Poujoulet}, {Thuillot}, {Walmsley}, {Brown},
  {Vallenari}, {Prusti}, {de Bruijne}, {Babusiaux}, {Bailer-Jones}, {Biermann},
  {Evans}, {Eyer}, {Jansen}, {Jordi}, {Klioner}, {Lammers}, {Lindegren},
  {Luri}, {Panem}, {Pourbaix}, {Randich}, {Sartoretti}, {Siddiqui}, {Soubiran},
  {van Leeuwen}, {Walton}, {Arenou}, {Bastian}, {Cropper}, {Drimmel}, {Katz},
  {Lattanzi}, {Bakker}, {Cacciari}, {Casta{\~n}eda}, {Chaoul}, {Cheek},
  {Fabricius}, {Guerra}, {Holl}, {Masana}, {Messineo}, {Mowlavi},
  {Nienartowicz}, {Panuzzo}, {Riello}, {Seabroke}, {Th{\'e}venin},
  {Gracia-Abril}, {Comoretto}, {Garcia-Reinaldos}, {Teyssier}, {Altmann},
  {Andrae}, {Audard}, {Bellas-Velidis}, {Benson}, {Blomme}, {Burgess}, {Busso},
  {Clementini}, {Clotet}, {Creevey}, {Davidson}, {De Ridder}, {Delchambre},
  {Ducourant}, {Fern{\'a}ndez-Hern{\'a}ndez}, {Fouesneau}, {Fr{\'e}mat},
  {Garc{\'\i}a-Torres}, {Gonz{\'a}lez-N{\'u}{\~n}ez}, {Gonz{\'a}lez-Vidal},
  {Gosset}, {Guy}, {Halbwachs}, {Hambly}, {Harrison}, {Hodgkin}, {Hutton},
  {Jasniewicz}, {Jean-Antoine-Piccolo}, {Jordan}, {Korn}, {Krone-Martins},
  {Lanzafame}, {Lebzelter}, {L{\"o}}, {Manteiga}, {Marrese},
  {Mart{\'\i}n-Fleitas}, {Moitinho}, {Mora}, {Osinde}, {Pancino},
  {Recio-Blanco}, {Richards}, {Rimoldini}, {Robin}, {Sarro}, {Siopis}, {Smith},
  {Sozzetti}, {S{\"u}veges}, {Torra}, {van Reeven}, {Abbas}, {Abreu Aramburu},
  {Accart}, {Aerts}, {Altavilla}, {{\'A}lvarez}, {Alvarez}, {Alves},
  {Anderson}, {Andrei}, {Anglada Varela}, {Antiche}, {Antoja}, {Arcay},
  {Astraatmadja}, {Bach}, {Baker}, {Balaguer-N{\'u}{\~n}ez}, {Balm}, {Barache},
  {Barata}, {Barbato}, {Barblan}, {Barklem}, {Barrado}, {Barros}, {Barstow},
  {Bartholom{\'e} Mu{\~n}oz}, {Bassilana}, {Becciani}, {Bellazzini},
  {Berihuete}, {Bertone}, {Bianchi}, {Bienaym{\'e}}, {Blanco-Cuaresma}, {Boch},
  {Boeche}, {Bombrun}, {Borrachero}, {Bossini}, {Bouquillon}, {Bourda},
  {Bragaglia}, {Bramante}, {Breddels}, {Bressan}, {Brouillet},
  {Br{\"u}semeister}, {Brugaletta}, {Bucciarelli}, {Burlacu}, {Busonero},
  {Butkevich}, {Buzzi}, {Caffau}, {Cancelliere}, {Cannizzaro}, {Cantat-Gaudin},
  {Carballo}, {Carlucci}, {Carrasco}, {Casamiquela}, {Castellani},
  {Castro-Ginard}, {Charlot}, {Chemin}, {Chiavassa}, {Cocozza}, {Costigan},
  {Cowell}, {Crifo}, {Crosta}, {Crowley}, {Cuypers}, {Dafonte}, {Damerdji},
  {Dapergolas}, {David}, {de Laverny}, {De Luise}, {De March}, {de Souza}, {de
  Torres}, {Debosscher}, {del Pozo}, {Delgado}, {Delgado}, {Diakite}, {Diener},
  {Distefano}, {Dolding}, {Drazinos}, {Dur{\'a}n}, {Edvardsson}, {Enke},
  {Eriksson}, {Esquej}, {Eynard Bontemps}, {Fabre}, {Fabrizio}, {Faigler},
  {Falc{\~a}o}, {Farr{\`a}s Casas}, {Federici}, {Fedorets}, {Fernique},
  {Figueras}, {Filippi}, {Findeisen}, {Fonti}, {Fraile}, {Fraser}, {Gai},
  {Galleti}, {Garabato}, {Garc{\'\i}a-Sedano}, {Garofalo}, {Garralda}, {Gavel},
  {Gavras}, {Gerssen}, {Geyer}, {Giacobbe}, {Gilmore}, {Girona}, {Giuffrida},
  {Glass}, {Gomes}, {Gueguen}, {Guerrier}, {Guti{\'e}}, {Haigron},
  {Hatzidimitriou}, {Hauser}, {Haywood}, {Heiter}, {Helmi}, {Heu}, {Hilger},
  {Hobbs}, {Hofmann}, {Holland}, {Huckle}, {Hypki}, {Icardi}, {Jan{\ss}en},
  {Jevardat de Fombelle}, {Jonker}, {Juh{\'a}sz}, {Julbe}, {Karampelas},
  {Kewley}, {Klar}, {Kochoska}, {Kohley}, {Kolenberg}, {Kontizas}, {Kontizas},
  {Koposov}, {Kordopatis}, {Kostrzewa-Rutkowska}, {Koubsky}, {Lambert},
  {Lanza}, {Lasne}, {Lavigne}, {Le Fustec}, {Le Poncin-Lafitte}, {Lebreton},
  {Leccia}, {Leclerc}, {Lecoeur-Taibi}, {Lenhardt}, {Leroux}, {Liao}, {Licata},
  {Lindstr{\o}m}, {Lister}, {Livanou}, {Lobel}, {L{\'o}pez}, {Managau}, {Mann},
  {Mantelet}, {Marchal}, {Marchant}, {Marconi}, {Marinoni}, {Marschalk{\'o}},
  {Marshall}, {Martino}, {Marton}, {Mary}, {Massari}, {Matijevi{\v{c}}},
  {Mazeh}, {McMillan}, {Messina}, {Michalik}, {Millar}, {Molina}, {Molinaro},
  {Moln{\'a}r}, {Montegriffo}, {Mor}, {Morbidelli}, {Morel}, {Morris},
  {Mulone}, {Muraveva}, {Musella}, {Nelemans}, {Nicastro}, {Noval},
  {O'Mullane}, {Ord{\'o}{\~n}ez-Blanco}, {Osborne}, {Pagani}, {Pagano},
  {Pailler}, {Palacin}, {Palaversa}, {Panahi}, {Pawlak}, {Piersimoni},
  {Pineau}, {Plachy}, {Plum}, {Poggio}, {Pr{\v{s}}a}, {Pulone}, {Racero},
  {Ragaini}, {Rambaux}, {Ramos-Lerate}, {Regibo}, {Reyl{\'e}}, {Riclet},
  {Ripepi}, {Riva}, {Rivard}, {Rixon}, {Roegiers}, {Roelens},
  {Romero-G{\'o}mez}, {Rowell}, {Royer}, {Ruiz-Dern}, {Sadowski}, {Sagrist{\`a}
  Sell{\'e}s}, {Sahlmann}, {Salgado}, {Salguero}, {Sanna}, {Santana-Ros},
  {Sarasso}, {Savietto}, {Schultheis}, {Sciacca}, {Segol}, {Segovia},
  {S{\'e}gransan}, {Shih}, {Siltala}, {Silva}, {Smart}, {Smith}, {Solano},
  {Solitro}, {Sordo}, {Soria Nieto}, {Souchay}, {Spagna}, {Stampa}, {Steele},
  {Steidelm{\"u}ller}, {Stephenson}, {Stoev}, {Suess}, {Surdej}, {Szabados},
  {Szegedi-Elek}, {Tapiador}, {Taris}, {Tauran}, {Taylor}, {Teixeira},
  {Terrett}, {Teyssandier}, {Titarenko}, {Torra Clotet}, {Turon}, {Ulla},
  {Utrilla}, {Uzzi}, {Vaillant}, {Valentini}, {Valette}, {van Elteren}, {Van
  Hemelryck}, {van Leeuwen}, {Vaschetto}, {Vecchiato}, {Veljanoski}, {Viala},
  {Vicente}, {Vogt}, {von Essen}, {Voss}, {Votruba}, {Voutsinas}, {Weiler},
  {Wertz}, {Wevers}, {Wyrzykowski}, {Yoldas}, {{\v{Z}}erjal}, {Ziaeepour},
  {Zorec}, {Zschocke}, {Zucker}, {Zurbach}, \& {Zwitter}}]{gaia2018}
{Gaia Collaboration}, {Spoto}, F., {Tanga}, P., {et~al.} 2018, \aap, 616, A13

\bibitem[{{Hanu{\v s}} {et~al.}(2013){Hanu{\v s}}, {Bro{\v z}}, {\v{D}urech},
  {Warner}, {Brinsfield}, {Durkee}, {Higgins}, {Koff}, {Oey}, {Pilcher},
  {Stephens}, {Strabla}, {Ulisse}, \& {Girelli}}]{hetal2013}
{Hanu{\v s}}, J., {Bro{\v z}}, M., {\v{D}urech}, J., {et~al.} 2013, \aap, 559,
  A134

\bibitem[{{Hanu{\v s}} {et~al.}(2016){Hanu{\v s}}, {{\v D}urech}, {Oszkiewicz},
  {Behrend}, {Carry}, {Delbo}, {Adam}, {Afonina}, {Anquetin}, {Antonini},
  {Arnold}, {Audejean}, {Aurard}, {Bachschmidt}, {Baduel}, {Barbotin},
  {Barroy}, {Baudouin}, {Berard}, {Berger}, {Bernasconi}, {Bosch}, {Bouley},
  {Bozhinova}, {Brinsfield}, {Brunetto}, {Canaud}, {Caron}, {Carrier},
  {Casalnuovo}, {Casulli}, {Cerda}, {Chalamet}, {Charbonnel}, {Chinaglia},
  {Cikota}, {Colas}, {Coliac}, {Collet}, {Coloma}, {Conjat}, {Conseil},
  {Costa}, {Crippa}, {Cristofanelli}, {Damerdji}, {Deback{\`e}re}, {Decock},
  {D{\'e}hais}, {D{\'e}l{\'e}age}, {Delmelle}, {Demeautis},
  {Dr{\'o}{\.z}d{\.z}}, {Dubos}, {Dulcamara}, {Dumont}, {Durkee}, {Dymock},
  {Escalante del Valle}, {Esseiva}, {Esseiva}, {Esteban}, {Fauchez},
  {Fauerbach}, {Fauvaud}, {Fauvaud}, {Forn{\'e}}, {Fournel}, {Fradet},
  {Garlitz}, {Gerteis}, {Gillier}, {Gillon}, {Giraud}, {Godard}, {Goncalves},
  {Hamanowa}, {Hamanowa}, {Hay}, {Hellmich}, {Heterier}, {Higgins}, {Hirsch},
  {Hodosan}, {Hren}, {Hygate}, {Innocent}, {Jacquinot}, {Jawahar}, {Jehin},
  {Jerosimic}, {Klotz}, {Koff}, {Korlevic}, {Kosturkiewicz}, {Krafft},
  {Krugly}, {Kugel}, {Labrevoir}, {Lecacheux}, {Lehk{\'y}}, {Leroy},
  {Lesquerbault}, {Lopez-Gonzales}, {Lutz}, {Mallecot}, {Manfroid}, {Manzini},
  {Marciniak}, {Martin}, {Modave}, {Montaigut}, {Montier}, {Morelle}, {Morton},
  {Mottola}, {Naves}, {Nomen}, {Oey}, {Og{\l}oza}, {Paiella}, {Pallares},
  {Peyrot}, {Pilcher}, {Pirenne}, {Piron}, {Poli{\'n}ska}, {Polotto}, {Poncy},
  {Previt}, {Reignier}, {Renauld}, {Ricci}, {Richard}, {Rinner}, {Risoldi},
  {Robilliard}, {Romeuf}, {Rousseau}, {Roy}, {Ruthroff}, {Salom}, {Salvador},
  {Sanchez}, {Santana-Ros}, {Scholz}, {S{\'e}n{\'e}}, {Skiff}, {Sobkowiak},
  {Sogorb}, {Sold{\'a}n}, {Spiridakis}, {Splanska}, {Sposetti}, {Starkey},
  {Stephens}, {Stiepen}, {Stoss}, {Strajnic}, {Teng}, {Tumolo}, {Vagnozzi},
  {Vanoutryve}, {Vugnon}, {Warner}, {Waucomont}, {Wertz}, {Winiarski}, \&
  {Wolf}}]{hetal2016}
{Hanu{\v s}}, J., {{\v D}urech}, J., {Oszkiewicz}, D.~A., {et~al.} 2016, \aap,
  586, A108

\bibitem[{{Haponiak} {et~al.}(2020){Haponiak}, {Breiter}, \&
  {Vokrouhlick{\'y}}}]{hap2020}
{Haponiak}, J., {Breiter}, S., \& {Vokrouhlick{\'y}}, D. 2020, Celestial
  Mechanics and Dynamical Astronomy, 132, 24

\bibitem[{{Henrard}(1982)}]{h1982}
{Henrard}, J. 1982, Celestial Mechanics, 27, 3

\bibitem[{{Henrard} \& {Murigande}(1987)}]{hm1987}
{Henrard}, J. \& {Murigande}, C. 1987, Celestial Mechanics, 40, 345

\bibitem[{{Herald} {et~al.}(2020){Herald}, {Gault}, {Anderson}, {Dunham},
  {Frappa}, {Hayamizu}, {Kerr}, {Miyashita}, {Moore}, {Pavlov}, {Preston},
  {Talbot}, \& {Timerson}}]{hetal2020}
{Herald}, D., {Gault}, D., {Anderson}, R., {et~al.} 2020, \mnras, 499, 4570

\bibitem[{{Hirayama}(1918)}]{hira18}
{Hirayama}, K. 1918, \aj, 31, 185

\bibitem[{{Jehin} {et~al.}(2011){Jehin}, {Gillon}, {Queloz}, {Magain},
  {Manfroid}, {Chantry}, {Lendl}, {Hutsem{\'e}kers}, \& {Udry}}]{jeh2011}
{Jehin}, E., {Gillon}, M., {Queloz}, D., {et~al.} 2011, The Messenger, 145, 2

\bibitem[{{Kaasalainen} \& {Lamberg}(2006)}]{kl2006}
{Kaasalainen}, M. \& {Lamberg}, L. 2006, Inverse Problems, 22, 749

\bibitem[{{Kaasalainen} {et~al.}(2001){Kaasalainen}, {Torppa}, \&
  {Muinonen}}]{kaa2001}
{Kaasalainen}, M., {Torppa}, J., \& {Muinonen}, K. 2001, \icarus, 153, 37

\bibitem[{{Kochanek} {et~al.}(2017){Kochanek}, {Shappee}, {Stanek}, {Holoien},
  {Thompson}, {Prieto}, {Dong}, {Shields}, {Will}, {Britt}, {Perzanowski}, \&
  {Pojma{\'n}ski}}]{Kochanek2017}
{Kochanek}, C.~S., {Shappee}, B.~J., {Stanek}, K.~Z., {et~al.} 2017, \pasp,
  129, 104502

\bibitem[{{Laskar}(1988)}]{l1988}
{Laskar}, J. 1988, \aap, 198, 341

\bibitem[{{Masiero} {et~al.}(2013){Masiero}, {Mainzer}, {Bauer}, {Grav},
  {Nugent}, \& {Stevenson}}]{metal13}
{Masiero}, J.~R., {Mainzer}, A.~K., {Bauer}, J.~M., {et~al.} 2013, \apj, 770, 7

\bibitem[{{Masiero} {et~al.}(2011){Masiero}, {Mainzer}, {Grav}, {Bauer},
  {Cutri}, {Dailey}, {Eisenhardt}, {McMillan}, {Spahr}, {Skrutskie}, {Tholen},
  {Walker}, {Wright}, {DeBaun}, {Elsbury}, {Gautier}, {Gomillion}, \&
  {Wilkins}}]{metal11}
{Masiero}, J.~R., {Mainzer}, A.~K., {Grav}, T., {et~al.} 2011, \apj, 741, 68

\bibitem[{{Nesvorn{\'y}} {et~al.}(2015){Nesvorn{\'y}}, {Bro{\v z}}, \&
  {Carruba}}]{netal15}
{Nesvorn{\'y}}, D., {Bro{\v z}}, M., \& {Carruba}, V. 2015, in Asteroids IV,
  ed. P.~{Michel}, F.~E. {DeMeo}, \& W.~F. {Bottke}, 297--321

\bibitem[{{Nesvorn{\'y}} \& {Vokrouhlick{\'y}}(2007)}]{nv2007}
{Nesvorn{\'y}}, D. \& {Vokrouhlick{\'y}}, D. 2007, \aj, 134, 1750

\bibitem[{{Nesvorn{\'y}} \& {Vokrouhlick{\'y}}(2008)}]{nv2008}
{Nesvorn{\'y}}, D. \& {Vokrouhlick{\'y}}, D. 2008, \aj, 136, 291

\bibitem[{{Rubincam}(2000)}]{rub2000}
{Rubincam}, D.~P. 2000, \icarus, 148, 2

\bibitem[{{Saillenfest} {et~al.}(2019){Saillenfest}, {Laskar}, \&
  {Bou{\'e}}}]{sail2019}
{Saillenfest}, M., {Laskar}, J., \& {Bou{\'e}}, G. 2019, \aap, 623, A4

\bibitem[{{Shappee} {et~al.}(2014){Shappee}, {Prieto}, {Stanek}, {Kochanek},
  {Holoien}, {Jencson}, {Basu}, {Beacom}, {Szczygiel}, {Pojmanski},
  {Brimacombe}, {Dubberley}, {Elphick}, {Foale}, {Hawkins}, {Mullins},
  {Rosing}, {Ross}, \& {Walker}}]{Shappee2014b}
{Shappee}, B., {Prieto}, J., {Stanek}, K.~Z., {et~al.} 2014, in American
  Astronomical Society Meeting Abstracts, Vol. 223, American Astronomical
  Society Meeting Abstracts \#223, 236.03

\bibitem[{{Slivan}(2002)}]{slivan02}
{Slivan}, S.~M. 2002, \nat, 419, 49

\bibitem[{{Slivan} \& {Binzel}(1996)}]{sb1996}
{Slivan}, S.~M. \& {Binzel}, R.~P. 1996, \icarus, 124, 452

\bibitem[{{Slivan} {et~al.}(2008){Slivan}, {Binzel}, {Boroumand }, {Pan},
  {Simpson}, {Tanabe}, {Villastrigo}, {Yen}, {Ditteon}, {Pray}, \&
  {Stephens}}]{setal2008}
{Slivan}, S.~M., {Binzel}, R.~P., {Boroumand }, S.~C., {et~al.} 2008, \icarus,
  195, 226

\bibitem[{{Slivan} {et~al.}(2003){Slivan}, {Binzel}, {Crespo da Silva},
  {Kaasalainen}, {Lyndaker}, \& {Kr{\v{c}}o}}]{setal2003}
{Slivan}, S.~M., {Binzel}, R.~P., {Crespo da Silva}, L.~D., {et~al.} 2003,
  \icarus, 162, 285

\bibitem[{{Slivan} {et~al.}(2009){Slivan}, {Binzel}, {Kaasalainen}, {Hock},
  {Klesman}, {Eckelman}, \& {Stephens}}]{setal2009}
{Slivan}, S.~M., {Binzel}, R.~P., {Kaasalainen}, M., {et~al.} 2009, \icarus,
  200, 514

\bibitem[{{Slivan} \& {Molnar}(2012)}]{karin2012}
{Slivan}, S.~M. \& {Molnar}, L.~A. 2012, \icarus, 220, 1097

\bibitem[{{Stephens}(2014)}]{s2014}
{Stephens}, R.~D. 2014, Minor Planet Bulletin, 41, 13

\bibitem[{{Szak{\'a}ts} {et~al.}(2020){Szak{\'a}ts}, {M{\"u}ller},
  {Al{\'\i}-Lagoa}, {Marton}, {Farkas-Tak{\'a}cs}, {B{\'a}nyai}, \&
  {Kiss}}]{setal2020}
{Szak{\'a}ts}, R., {M{\"u}ller}, T., {Al{\'\i}-Lagoa}, V., {et~al.} 2020, \aap,
  635, A54

\bibitem[{{Tonry} {et~al.}(2018){Tonry}, {Denneau}, {Heinze}, {Stalder},
  {Smith}, {Smartt}, {Stubbs}, {Weiland}, \& {Rest}}]{tetal2018}
{Tonry}, J.~L., {Denneau}, L., {Heinze}, A.~N., {et~al.} 2018, \pasp, 130,
  064505

\bibitem[{{Vokrouhlick\'y} {et~al.}(2015){Vokrouhlick\'y}, {Bottke}, {Chesley},
  {Scheeres}, \& {Statler}}]{vetal15}
{Vokrouhlick\'y}, D., {Bottke}, W.~F., {Chesley}, S.~R., {Scheeres}, D.~J., \&
  {Statler}, T.~S. 2015, in Asteroids IV, ed. P.~{Michel}, F.~E. {DeMeo}, \&
  W.~F. {Bottke}, 509--531

\bibitem[{{Vokrouhlick{\'y}} {et~al.}(2003){Vokrouhlick{\'y}}, {Nesvorn{\'y}},
  \& {Bottke}}]{vetal2003}
{Vokrouhlick{\'y}}, D., {Nesvorn{\'y}}, D., \& {Bottke}, W.~F. 2003, \nat, 425,
  147

\bibitem[{{Vokrouhlick{\'y}} {et~al.}(2006){Vokrouhlick{\'y}}, {Nesvorn{\'y}},
  \& {Bottke}}]{vetal2006c}
{Vokrouhlick{\'y}}, D., {Nesvorn{\'y}}, D., \& {Bottke}, W.~F. 2006, \icarus,
  184, 1

\bibitem[{{Vokrouhlick{\'y}} \& {{\v C}apek}(2002)}]{vc2002}
{Vokrouhlick{\'y}}, D. \& {{\v C}apek}, D. 2002, \icarus, 159, 449

\bibitem[{{Ward} \& {Hamilton}(2004)}]{wh2004}
{Ward}, W.~R. \& {Hamilton}, D.~P. 2004, \aj, 128, 2501

\bibitem[{{Warner} {et~al.}(2009){Warner}, {Harris}, \& {Pravec}}]{lcdb}
{Warner}, B.~D., {Harris}, A.~W., \& {Pravec}, P. 2009, \icarus, 202, 134

\end{thebibliography}
\end{document}